\documentclass[12pt]{article}
\usepackage{citesort}
\usepackage[dvips]{graphicx}

\newcommand{\third}{\frac{1}{3}}
\newcommand{\be}{\begin{equation}}
\newcommand{\ee}{\end{equation}}
\newcommand{\bea}{\begin{eqnarray}}
\newcommand{\eea}{\end{eqnarray}}
\renewcommand{\theequation}{\arabic{section}.\arabic{equation}}

\begin{document}
\begin{titlepage}
 
\vspace{1in}
 
\begin{center}
\Large
{\bf Qualitative Analysis of Isotropic Curvature String Cosmologies}
 
\vspace{1in}

\normalsize
 
\large{Andrew P. Billyard$^{1a}$, Alan A. Coley$^{1,2b}$
 \& James E. Lidsey$^{3c}$}

\normalsize
\vspace{.7in}
 
{\em $^1$Department of Physics, \\ Dalhousie University, Halifax, NS, B3H 3J5, 
Canada} \\

\vspace{.1in}

{\em $^2$Department of Mathematics and Statistics, \\ Dalhousie University, 
Halifax, NS, B3H 3J5, 
Canada} \\

\vspace{.1in}

{\em $^3$Astronomy Unit, School of Mathematical Sciences, \\
Queen Mary and Westfield, Mile End Road, London, E1 4NS, U. K.}

\end{center}
 
\vspace{.7in}
 
\baselineskip=24pt
\begin{abstract}
A complete qualitative study of the dynamics of string cosmologies is
presented for the class of isotopic curvature universes. These models
are of Bianchi types I, V and IX and reduce to the general class of
Friedmann--Robertson--Walker universes in the limit of vanishing shear
isotropy.  A non--trivial two--form potential and cosmological
constant terms are included in the system.  In general, the two--form
potential and spatial curvature terms are only dynamically important
at intermediate stages of the evolution. In many of the models, the
cosmological constant is important asymptotically and anisotropy
becomes dynamically negligible. There also exist bouncing cosmologies.

\end{abstract}

PACS NUMBERS: 98.80.Cq, 04.50+h, 98.80.Hw
 
\vspace{.3in}
$^a$Electronic mail: jaf@mscs.dal.ca

$^b$Electronic mail: aac@mscs.dal.ca

$^c$Electronic mail: jel@maths.qmw.ac.uk

\end{titlepage}

\setcounter{equation}{0}
\section{Introduction}

One of the strongest constraints that a unified theory 
of the fundamental interactions must 
satisfy is that it leads to realistic cosmological models. There 
are known to be five consistent perturbative superstring theories 
in ten dimensions. (See, e.g., Ref. \cite{GreSchWit87}). 
The type II and 
heterotic theories each contain a Neveu--Schwarz/Neveu--Schwarz
(NS--NS) sector of bosonic massless excitations that includes 
a scalar dilaton field, a graviton and an antisymmetric two--form 
potential. The interactions between these fields lead to significant 
deviations from the standard, hot big bang model 
based on conventional Einstein gravity \cite{Veneziano91}
and a study of the cosmological consequences of superstring theory is 
therefore important. 

In string cosmology, the dynamics of the universe below the string scale
is determined by the effective supergravity actions. The NS--NS string 
cosmologies contain the non--trivial fields discussed above.
Recently \cite{BillyardColeyLidsey1} (hereafter referred to as paper 
I), a complete qualitative analysis for the 
spatially flat Friedmann--Robertson--Walker (FRW) and 
axisymmetric Bianchi type I NS--NS string cosmologies was presented.
A central charge deficit was also included 
and found to have significant effects on the nature of 
the equilibrium points. This study unified and extended previous 
qualitative analyses of this system \cite{unifyprevious,k,emw}.

In Ref. \cite{BillyardColeyLidsey2} 
(hereafter referred to as paper II), a phenomenological 
cosmological constant was introduced into the string frame effective 
action in such a way that it was not coupled directly 
to the dilaton field. 
Such a term yields valuable insight into the dynamics
of more general string models containing non--trivial Ramond--Ramond (RR) 
fields. The interplay between such a 
cosmological constant and the two--form potential had not been considered
previously. It was found that the interactions led to
heteroclinic orbits in the phase space, where 
the universe underwent a series of
oscillations between expanding and contracting phases. 

The purpose of the present paper is to extend the work of
\cite{BillyardColeyLidsey1,BillyardColeyLidsey2} 
to both isotropic and anisotropic  
cosmologies containing spatial curvature terms. In particular, 
we consider the class of `isotropic curvature' universes \cite{MacCallum,Mac}. 
These are spatially homogeneous but contain non-trivial curvature and 
anisotropy. The characteristic feature of these models is that 
the three-dimensional Ricci curvature tensor is isotropic. In other words,  
${^{(3)}}R_{ij}$ is proportional to $k\delta_{ij}$ on the spatial
hypersurfaces and these surfaces therefore have 
constant curvature $k$ \cite{MacCallum}. 
The class of isotropic curvature universes contains the Bianchi type I ($k=0$)
and V ($k<0$) models and a special case of the Bianchi type IX ($k>0$)
models \cite{Mac}.  

The four--dimensional line element of these cosmologies is given by 
\cite{MacCallum,Mac}
\begin{equation}
ds^2 =-dt^2 +e^{2\alpha (t)} \left[ \left(\mathbf{\omega}^1\right)^2
	+e^{-2\sqrt3\beta_s(t)}\left(\mathbf{\omega}^2\right)^2 +
	e^{2 \sqrt3\beta_s(t)} \left(\mathbf{\omega}^3\right)^2\right], 
	\label{metric}
\end{equation}
where the one-forms $\{\omega^1, \omega^2, \omega^3\}$ are given in \cite{MacCallum,Mac}.  The spatial hypersurfaces, $t={\rm constant}$, are the surfaces 
of homogeneity. The variable, $\alpha$, parametrizes 
the effective spatial volume of the universe and $\beta$ 
determines the level of anisotropy. We refer to it as the shear parameter. 
In essence, the isotropic curvature models can be regarded
as the simplest anisotropic generalizations of the 
flat $(k=0)$, open $(k<0)$ and closed $(k>0)$ FRW universes, respectively. 
The isotropic models are recovered when $\dot{\beta}=0$.

The paper is organized as follows.  In Section 2, we summarize 
the form of the effective actions we consider. 
The cosmological field equations are derived and the 
general asymptotic behaviour 
of the two--form potential and spatial curvature is discussed. 
The global qualitative dynamics for the different classes of models 
is determined in Sections 3 and 4. 
We summarize and conclude our results in Section 5. 
The functional forms of the
equilibrium points that arise in the analysis are presented in the 
Appendix. 

\section{Cosmological Field Equations}

\setcounter{equation}{0}

\subsection{String Effective Action}

The NS--NS sector of massless bosonic excitations is common to both the 
type II and heterotic superstring theories. 
The four--dimensional, string effective action for the NS--NS 
fields can be written as \cite{effective}
\begin{equation}
\label{NSaction}
S=\int d^4 x  \sqrt{-g} e^{- \Phi} \left[ R +\left( \nabla \Phi \right)^2 - 
\frac{1}{12} H_{\mu\nu\lambda}H^{\mu\nu\lambda} -2\Lambda \right]  ,
\end{equation}
where the string coupling, $g_s^2 \equiv e^{\Phi}$, is 
determined by the value of the dilaton field, $\Phi$, 
the space--time manifold has metric, $g_{\mu\nu}$, and 
Ricci curvature, $R$, the antisymmetric two--form, $B_{\mu\nu}$,
has a field strength $H_{\mu\nu\lambda} 
\equiv \partial_{[\mu} B_{\nu\lambda ]}$, and 
$g\equiv {\rm det}g_{\mu\nu}$. 
The central charge deficit of the theory is denoted by 
the constant, $\Lambda$. The value of this term depends on 
the conformal field theory that is coupled to the string and it 
can be positive or negative. 
Such a term may also arise from the compactification 
of higher--dimensional form fields or non--perturbative corrections 
to the self--interaction of the dilaton field \cite{kko98}.

The three--form $H_{\mu\nu\lambda}$ is 
dual to a one--form in four dimensions and the field equation for
$B_{\mu\nu}$ is solved by the ansatz \cite{sen}
\begin{equation}
\label{sigma}
H^{\mu\nu\lambda} \equiv e^{\Phi} \epsilon^{\mu\nu\lambda\kappa}
\nabla_{\kappa} \sigma   ,
\end{equation}
where $\epsilon^{\mu\nu\lambda\kappa}$ is the covariantly 
constant four--form. The scalar $\sigma$ may be 
interpreted as a pseudo--scalar `axion' field. The Bianchi identity, 
$\partial_{[ \mu}H_{\nu\lambda\kappa ]} \equiv 0$, is then 
solved by reinterpreting this constraint as the field equation for 
$\sigma$ and this latter equation 
can be derived from the dual effective action \cite{sen}
\begin{equation}
\label{sigmaaction}
S=\int d^4 x \sqrt{-g} e^{-\Phi} \left[ R +\left( \nabla \Phi \right)^2 
-\frac{1}{2}e^{2\Phi} \left( \nabla \sigma \right)^2 -2\Lambda \right] .
\end{equation}
We establish the dynamics of cosmological models derived from Eq. 
(\ref{sigmaaction}) in Section 3. 

In paper II, the action 
\begin{equation}
\label{massivefourA}
S=\int d^4 x \sqrt{-g} \left\{ e^{-\Phi} \left[ R + \left( \nabla 
\Phi \right)^2 -\frac{1}{2} e^{2\Phi} \left( \nabla \sigma \right)^2 
\right] -\Lambda_{\rm M} \right\}
\end{equation}
was considered, 
where $\Lambda_{\rm M}$ was interpreted as a phenomenological 
cosmological constant arising from the interaction potential
of a slowly  rolling scalar field. We now proceed to discuss 
this action within the context of 
the massive type IIA supergravity theory in ten 
dimensions \cite{Romans}. 
This theory represents the low--energy limit of the type 
IIA superstring and 
has been the subject of renewed interest recently following the advances 
that have been made 
in our understanding of the non--perturbative features of string theory 
\cite{tow,massivepapers}.  

In this theory, the NS--NS two--form potential becomes massive. In the 
absence of such a field, the action is given in the string frame by \cite{tow}
\begin{equation}
\label{massiveten}
S=\int d^{10} x \sqrt{-g_{10}} \left\{ e^{-\varsigma} \left[ R_{10} +\left( 
\nabla \varsigma \right)^2 \right] 
-\frac{1}{4} F_{AB}F^{AB}  - \frac{1}{48} 
F_{ABCD}F^{ABCD} -\frac{1}{2} m^2 \right\}  ,
\end{equation}
where $\varsigma$ represents the ten--dimensional 
dilaton field, $m^2$ is the cosmological constant 
and $A=(0, 1, \ldots , 9)$, etc. The 
antisymmetric field strengths $F_{AB}$ and $F_{ABCD}$ for the one-form 
and three--form potentials represent RR degrees of freedom 
because they do not couple directly to the dilaton field 
\cite{GreSchWit87}. The massless theory is recovered when $m=0$. 

Maharana and Singh \cite{MahSing} have employed the Kaluza--Klein 
technique \cite{SchSch,MahSch} to compactify the ten--dimensional 
theory (\ref{massiveten}) on a six--dimensional torus. We 
consider a truncation of the dimensionally reduced four--dimensional 
action, where we include only the components of the four--form 
on the four--dimensional external 
spacetime. We therefore neglect the moduli 
fields arising from the compactification of the form fields
and the gauge fields originating from the higher--dimensional metric. 
The effective four--dimensional action is then given by 
\begin{equation}
\label{massivefour}
S=\int d^4 x \sqrt{-g} \left\{ e^{-\Phi} \left[ R +\left( \nabla \Phi 
\right)^2 -6 \left( \nabla \gamma \right)^2  \right] 
 -\frac{1}{48} e^{6\gamma} F_{\mu\nu\lambda\kappa}
F^{\mu\nu\lambda\kappa} -\frac{1}{2} m^2 e^{6\gamma} \right\}   ,
\end{equation}
%\begin{eqnarray}
%\label{massivefour}
%S=\int d^4 x \sqrt{-g} \left\{ e^{-\Phi} \left[ R +\left( \nabla \Phi 
%\right)^2 -6 \left( \nabla \gamma \right)^2  \right] \right. \nonumber \\
%\left. -\frac{1}{48} e^{6\gamma} F_{\mu\nu\lambda\kappa}
%F^{\mu\nu\lambda\kappa} -\frac{1}{2} m^2 e^{6\gamma} \right\}   ,
%\end{eqnarray}
where the four--dimensional dilaton, $\Phi$, is defined 
in terms of the ten--dimensional dilaton by $\Phi \equiv \varsigma -6\gamma$ 
and $\gamma$ parametrizes the volume of the torus. 

The Bianchi identity for the three--form potential 
is trivially satisfied in four dimensions and its field equation 
is solved by the ansatz $F^{\mu\nu\lambda\kappa} = Q 
e^{-6\gamma}\epsilon^{\mu\nu\lambda\kappa}$, where $Q$ is an 
arbitrary constant and $\epsilon^{\mu\nu\lambda\kappa}$ is the covariantly 
constant four--form. 
Applying this duality, together with the conformal 
transformation 
\begin{equation}
\label{conformal}
\tilde{g}_{\mu\nu} = \Omega^2 g_{\mu\nu} , \qquad \Omega^2 \equiv e^{-\Phi} 
\end{equation}
implies that the effective four--dimensional action may be 
expressed in the Einstein frame as
\begin{equation}
\label{Einsteinaction}
S=\int d^4 x \sqrt{-\tilde{g}} \left[ \tilde{R}  -\frac{1}{2} \left( 
\tilde{\nabla} \Phi \right)^2 -\frac{1}{2} \left( \tilde{\nabla} 
y \right)^2  
-\frac{1}{2} Q^2 e^{2\Phi -\sqrt{3}y} -\frac{1}{2} m^2 e^{2\Phi +\sqrt{3}y} 
\right] , 
\end{equation}
%\begin{eqnarray}
%\label{Einsteinaction}
%S=\int d^4 x \sqrt{-\tilde{g}} \left[ \tilde{R}  -\frac{1}{2} \left( 
%\tilde{\nabla} \Phi \right)^2 -\frac{1}{2} \left( \tilde{\nabla} 
%y \right)^2 \right. \nonumber \\
%\left. 
%-\frac{1}{2} Q^2 e^{2\Phi -\sqrt{3}y} -\frac{1}{2} m^2 e^{2\Phi +\sqrt{3}y} 
%\right] , 
%\end{eqnarray}
where $y \equiv \sqrt{12}\gamma$ represents a rescaled modulus 
field. 

Without loss of generality,  one may perform 
linear translations on the dilaton and modulus fields
such that the parameters $Q$ and 
$m$ become effectively equal. Thus, 
action (\ref{Einsteinaction}) may be written in the form
\begin{equation}
\label{coshaction}
S = \int d^4 x \sqrt{-\tilde{g}} \left[ \tilde{R} -\frac{1}{2} \left( 
\tilde{\nabla} \Phi \right)^2 -\frac{1}{2} \left( \tilde{\nabla} y 
\right)^2 -\Lambda_{\rm M}e^{2\Phi} \cosh \left( \sqrt{3} y \right) \right]  ,
\end{equation}
where $\Lambda_{\rm M}$ is a positive--definite constant. 
It follows, therefore, that 
the four--form and ten--dimensional mass parameter together 
provide an effective potential for the modulus field 
with a global minimum located at $y  =0$. Consequently, 
the internal space can become stabilized for this specific 
compactification and a consistent 
truncation of the effective action 
is therefore given by specifying 
$y =0$ in Eq. (\ref{coshaction}). Applying the inverse of the 
conformal transformation (\ref{conformal}) then
implies that the effective string frame action is given by Eq. 
(\ref{massivefourA}) when the axion field is trivial.  Thus, a truncated form 
of Eq. (\ref{massivefourA}) is relevant to the type IIA theory. 
The dynamics of these models is considered in Section 4. 

For the purposes of deriving the cosmological field equations 
from the effective actions (\ref{sigmaaction}) and 
(\ref{massivefourA}), we combine these two 
expressions into the single action
\begin{equation}
\label{singleaction}
S=\int d^4 x \sqrt{-g} \left\{ 
e^{-\Phi} \left[ R +\left( \nabla \Phi \right)^2 
-\frac{1}{2}e^{2\Phi} \left( \nabla \sigma \right)^2 -2\Lambda \right]
-\Lambda_{\rm M} \right\}
\end{equation}
where it is understood that either $\Lambda$ or $\Lambda_{\rm M}$ should be 
set to zero. We assume 
that all massless degrees of freedom are constant on the surfaces of 
homogeneity, $t={\rm constant}$. 
The field equations derived from action (\ref{singleaction}) 
for the isotropic curvature metric (\ref{metric}) are  then given by 
\begin{eqnarray}
\label{rr1}
\ddot{\alpha} -\dot{\alpha}\dot{\varphi} -\frac{1}{2} 
     \rho +\tilde K +\frac{1}{2}\Lambda_{\rm M}
     e^{\varphi+3\alpha} =0\\
\label{rr2}
2\ddot{\varphi} -\dot{\varphi}^2 -3\dot{\alpha}^2 - 6 \dot{\beta}^2 
     +\frac{1}{2}\rho - 3 \tilde K +2 \Lambda =0 \\
\label{rr3}
\ddot{\beta} -\dot{\beta} \dot{\varphi} =0 \\
\label{rr4}
\dot{\tilde K} +2\dot\alpha \tilde K = 0 \\
\label{rr5}
\dot{\rho} +6\dot{\alpha} \rho =0 
\end{eqnarray}
together with the generalized Friedmann constraint equation
\begin{equation}
\label{rrfriedmann}
3\dot{\alpha}^2 -\dot{\varphi}^2 +6 \dot{\beta}^2 +\frac{1}{2} \rho
-3\tilde K +2\Lambda + \Lambda_{\rm M} e^{\varphi+3\alpha} =0,
\end{equation}
where 
\begin{equation}
\label{varphi}
\varphi \equiv \Phi -3\alpha
\end{equation}
defines the `shifted' 
dilaton field \cite{Veneziano91,shifted}, 
\begin{equation}
\label{rho}
\rho \equiv \dot\sigma^2 e^{2\varphi+6\alpha }
\end{equation}
may be interpreted as the
effective  energy density of the pseudo--scalar axion field \cite{k},
\begin{equation}
\label{tildeK}
\tilde K\equiv 2k \exp(-2\alpha)
\end{equation}
represents the spatial curvature term, $\alpha$ and $\beta$ are defined
in Eq. (\ref{metric}), and a dot denotes differentiation with respect to 
cosmic time, $t$. 

The moduli fields that may also 
arise in the string effective action from 
the compactification of higher dimensions have not been included in Eq. 
(\ref{NSaction}) and it is assumed, in particular, 
that the internal dimensions 
are fixed. We emphasize, however, that in the NS--NS model
$(\Lambda_{\rm M} =0)$, one may readily include 
the dynamical effects of these extra dimensions in the case of a toroidal 
compactification, where the internal space has 
the topology $T=S^1 \times S^1 \times \ldots \times S^1$, 
by reinterpreting the shear parameter, $\beta$, in the field equations 
(\ref{rr1})--(\ref{rrfriedmann}) \cite{MahSch,BillyardColeyLidsey1}. 
Modulo a trivial rescaling, 
the moduli fields, $\beta_m$, that parametrize the radii of the circles, 
$S^1$, have the same functional form in the field 
equations as the shear term. 
These moduli may therefore be combined with the shear by replacing 
$\beta^2$ with an expression 
of the form $\Gamma^2 =\beta^2 + \sum \beta_m^2$. 

\subsection{Asymptotic Behaviour}

Before concluding this Section, we make some general remarks 
regarding the asymptotic behaviour of 
the cosmological models discussed above. 

It can be shown that the action (\ref{singleaction}) is invariant under
a global ${\rm SL}(2, R)$ transformation acting on the dilaton and axion
fields when the cosmological constants 
$\Lambda$ and $\Lambda_{\rm M}$ vanish \cite{sen}. The symmetry 
becomes manifest in the conformally related Einstein frame (\ref{conformal})
and may be employed to generate a non--trivial axion 
field from a solution where such a field is trivial \cite{clw}. 
Solutions containing a dynamical 
axion field are known as `dilaton--axion' cosmologies. The constant 
axion solutions are referred to as `dilaton--vacuum' solutions 
and are presented in Eq. (\ref{dmv}). 

The functional form of the dilaton--axion cosmologies 
has been derived \cite{clw}. 
The general feature exhibited by these models is that the axion field is 
dynamically important only for a short time interval. The solutions 
asymptotically approach one of the
dilaton--vacuum solutions (\ref{dmv}) in the high and low 
curvature regimes. The axion field also results in a lower bound on the 
value of the dilaton field and therefore the string coupling. 
However, the 
${\rm SL}(2, R)$ symmetry is broken when either of the 
cosmological constants is
present in the action (\ref{singleaction}) \cite{kms}
and analytical solutions are not 
known in this case, even in the isotropic limit $(\dot{\beta} =0)$. 

In addition, the variables $\tilde K$ and $\rho$ 
in Eqs. (\ref{rr4}) and 
(\ref{rr5}) may in general be combined to define the new variable
\begin{equation}
\Xi\equiv \frac{\rho-\tilde K}{\rho+\tilde K}  .
\end{equation}
This implies that Eqs. (\ref{rr4}) and (\ref{rr5}) 
are equivalent to the evolution equation
\begin{equation}
\dot\Xi = -2\dot\alpha\left(1-\Xi^2\right). \label{dchi}
\end{equation}
Hence, {\em all} of the equilibrium points occur either for $\dot\alpha=0$ or
for $\Xi^2=1$ and this implies that either $\rho=0$
($\Xi=-1$) or $\tilde K =0$ ($\Xi=+1$) asymptotically
if $\dot\alpha\neq0$.  

In general, Eqs.
(\ref{rr1})-(\ref{rrfriedmann}) define a four-dimensional dynamical
system. (Although there are five ODE's, Eq. (\ref{rrfriedmann})
may be employed to globally reduce the system by one dimension). In the
cases in which all of the equilibrium points lie on $\Xi^2=1$, the
asymptotic properties of the string cosmologies can be determined from
the dynamics in the three-dimensional sets $\rho=0$ or $\tilde K=0$. 
The latter three-dimensional dynamical system was studied in
papers I and II. 

In view of this, we explicitly examine the
three-dimensional $\rho=0$ case in what follows.  
The only case in which there exist
equilibrium points with $\dot\alpha=0$ but $\Xi^2\neq 1$ occurs in the NS-NS
case ($\Lambda_{\rm M}=0$) in which $\tilde K>0$ and $\Lambda>0$. 
We shall examine the full four-dimensional system in this
case, although the three-dimensional subset $\rho=0$ still plays a
principal r\^{o}le in the asymptotic analysis.

All solutions represented by the equilibrium points that arise in this
work are presented in the Appendix. Some of these points also arose in
papers I and II, where they were labelled differently.
In Table
\ref{table}, we list all the equilibrium points obtained and unify the
notation employed in the different works.

\begin{table}[ht]
\begin{center}
\begin{tabular}{|c|c|c|c|}
\hline
This Paper & Equation & Paper I \cite{BillyardColeyLidsey1} 
	& Paper II \cite{BillyardColeyLidsey2} \\
\hline \hline
$L^\pm$ ($S_1,S_2$)$^*$ & \ref{dmv} & $L_\pm$, $V$, ($R$, $A$, $S$, $S_1$, $S_2$)$^*$ & $W$, $W^\pm$, ($S_{u,v}$, $S_1$, $S_2$)$^*$ \\
\hline
$L_1$, ($C$)$^*$ & \ref{static_general} (\ref{static}) & ($C$)$^*$ & $-$ \\
\hline
$S_1^\pm$ & \ref{newsol} & $-$ & $F$ \\
\hline
$R$, $A$ & \ref{at_half} & $-$ & $R$, $A$ \\
\hline
$S^\pm$ & \ref{curv_drive} & $-$ & $-$  \\
\hline
 $N$ & \ref{open_new} & $-$ & $-$ \\
\hline
\end{tabular}
\end{center}
\caption{{\em The equilibrium points/sets obtained in this paper are
listed in the first column and their location in the  Appendix 
is given in the second column. The equivalent equilibrium points
(with different notation) in papers I \cite{BillyardColeyLidsey1} and II
\cite{BillyardColeyLidsey2} are listed in the appropriate row.
$^*$Note: The equilibrium points in parentheses in the first row are
just endpoints to the lines $L_\pm$, $V$ or $W$.  Similarly,
equilibrium point $C$ in the second row is the endpoint for the line
$L_1$ for $Y_1=Y_2=1$. (See the text for the definition 
of $Y_i$).  The points $R$ and $A$ in the first row
represent different solutions to those represented by the points $R$ 
and $A$ in the fourth
row.  Also, in the third row, point $F$ corresponds to point $S^-_1$
only.}}
\label{table}
\end{table}

\setcounter{equation}{0}
\section{Non-Zero Central Charge Deficit ($\Lambda_{\rm M}=0$)}

In this  Section we perform the qualitative analysis 
for the NS--NS string effective action (\ref{singleaction}) with 
$\Lambda_{\rm M} =0$ for an arbitrary central charge deficit. 
Through Eq. (\ref{rrfriedmann}), we eliminate the variable $\rho$
from the field equations, and make the following definitions:
\begin{equation}
X\equiv \frac{\sqrt3\dot\alpha}{\xi}, \quad
Y\equiv \frac{\dot\varphi}{\xi}, \quad
Z\equiv \frac{6\dot{\beta}^2}{\xi^2}, \quad
U\equiv \frac{\pm 3 \tilde K}{\xi^2}, \quad
V\equiv \frac{\pm 2\Lambda}{\xi^2}, \quad
\frac{d}{dt}\equiv \xi \frac{d}{dT} . \label{TheDefs}
\end{equation}
The $\pm$ signs in the definitions for $U$ and $V$ are to ensure that
$U>0$ and $V>0$ when necessary.  With these definitions, all
variables are bounded such that $0\leq\!\{X^2,Y^2,Z,U,V\}\leq\!1$ and Eq. 
(\ref{rrfriedmann}) now reads
\begin{equation}
\label{cf}
\frac{1}{2}\rho \xi^{-2} = Y^2 \pm U \mp V -X^2-Z \geq 0. \label{newFried1}
\end{equation}

The variable $\xi$ is defined in each of the following four cases 
(the section in which they occur is indicated in parentheses) by:
\begin{itemize}
\item $\Lambda>0$\begin{itemize} \item $\tilde K>0$: $\xi^2 \equiv 3\tilde K
	+\dot\varphi^2$ (Section 3.1),
\item $\tilde K<0$: $\xi^2 \equiv \dot\varphi^2$ (Section 3.2),\end{itemize}
\item $\Lambda<0$ \begin{itemize} \item $\tilde K>0$: $\xi^2 \equiv 3\tilde K
	+\dot\varphi^2-2\Lambda$ (Section 3.3),
\item $\tilde K<0$: $\xi^2 \equiv \dot\varphi^2-2\Lambda$ (Section 3.4).
\end{itemize}\end{itemize}
For example, consider $\Lambda>0$ with $\tilde K>0$; 
for this case $Y^2+U=1$ and Eq. (\ref{newFried1}) reads
\begin{eqnarray}
\frac{1}{2}\rho \xi^{-2} =1-V-X^2-Z \geq 0.
\end{eqnarray}
Hence, we use $\{X,V,Z\}$ as the phase space variables 
(see Section 3.1 for details).  We now consider each of these cases 
in turn and introduce 
subscripts $i=(1,2,3,4) $ to the variables $\{X,Y,Z,U,V\}$ to distinguish the 
different subsections. 

For each case, we will set up the four-dimensional dynamical system,
followed by a discussion of the $\tilde K=0$ invariant set as examined
in Paper I.  Note that in Section 3.2 the
$\tilde K=0$ case is identical to that presented in Section 3.1 and is
therefore omitted from that section.  Similarly, there will be no
$\tilde K=0$ discussion in Section 3.4 since it is discussed in
Section 3.3.  We will then examine the $\rho=0$ invariant set.  As
discussed in Section 2.2, all equilibrium points discussed below which
have $\dot\alpha=0$ also have $\Xi^2 \neq 1$, and hence nearly all the 
orbits asymptote towards the equilibrium points in one of the
invariant sets $\rho=0$ or $\tilde K=0$.  The qualitative behaviour of
the four-dimensional phase space in each case is then examined.

\subsection{The Case $\Lambda>0$, $\tilde K>0$}

We define $\xi^2 = \dot\varphi^2+3\tilde K$ and utilize the
positive signs for $U_1$ and $V_1$ as defined by Eq. 
(\ref{TheDefs}).  From the generalized Friedmann equation we have
that
\begin{equation}
0 \leq X_1^2 + Z_1 + V_1 \leq 1,\qquad Y_1^2+U_1=1,
\end{equation}
and therefore we may eliminate $U_1$ (which is proportional 
to  $\tilde K$), and consider the
four-dimensional system of ODEs for $0\leq \{X_1^2, Y_1^2, Z_1, V_1\}\leq1$:
\begin{eqnarray}
\label{dX_1}
\frac{dX_1}{dT} & = &\frac{1}{\sqrt3}\left(1-X_1^2\right)\left(2+Y_1^2\right)-\sqrt{3}\left(Z_1+V_1 \right) +X_1Y_1\left(1-X_1^2-Z_1\right), \\
\label{dY_1}
\frac{dY_1}{dT} &=& \left(1-Y_1^2\right) \left(X_1^2+Z_1
        +\frac{1}{\sqrt3}X_1Y_1\right),\\
\label{dZ_1}
\frac{dZ_1}{dT} &=& 2Z_1\left[ Y_1\left( 1-X_1^2-Z_1\right) 
        + \frac{1}{\sqrt3} X_1 \left( 1-Y_1^2 \right) \right], \\
\label{dV_1}
\frac{dV_1}{dT} &=& -2V_1 \left[ Y_1\left(X_1^2+Z_1\right) 
        -\frac{1}{\sqrt{3}} X_1	 \left( 1 - Y_1^2 \right) \right].
\end{eqnarray}
The invariant sets $Y_1^2=1$, $X_1^2+Z_1+V_1=1$,
$Z_1=0$ and $V_1=0$ define the boundary of the phase space.  The
equilibrium sets and their corresponding eigenvalues (denoted by
$\lambda$) are
\begin{eqnarray}
L^\pm: & & Y_1=\pm1, Z_1=1-X_1^2, V_1=0; \nonumber \\
   &&
   \left(\lambda_1,\lambda_2, \lambda_3, \lambda_4\right) =  \left(
    \mp\frac{2}{\sqrt3} \left[\sqrt3 \pm X_1 \right], \mp 2, 0,
	-2\sqrt3\left[X_1\pm\frac{1}{\sqrt3}\right]\right), \\
L_1: & & X_1=0,Z_1=0,V_1=\third(2+Y_1^2); \nonumber \\
   && 
   \left(\lambda_\pm, \lambda_2, \lambda_3\right) =  \left(
   \frac{1}{2} \left[ Y_1\pm\frac{1}{\sqrt3}\sqrt{19Y_1^2-16} \right ], 
   2Y_1, 0    \right). \label{mother}
\end{eqnarray}
The zero eigenvalues arise because these 
are all {\em lines} of equilibrium
points.  Here, the global sources are the lines $L_1$ (for $Y_1>0$ or
$\dot\varphi>0$) and $L^-$ (for $X_1<\frac{1}{\sqrt3}$).  The global sinks are
the lines $L_1$ (for $Y_1<0$) and $L^+$ (for $X_1>\frac{-1}{\sqrt3}$).

This case is {\em different} from the other three cases to be
considered in this Section, 
since it is the only one with the line of equilibrium
points, $L_1$, {\em inside} the phase space. This line acts as both sink
and source, and corresponds
to the exact static solution (\ref{static_general}) 
which generalizes the static `linear dilaton--vacuum'
solution (\ref{static}) \cite{myers}. 
This solution was examined in \cite{emw} for $\dot{\varphi} >0$ and was shown
by a perturbation analysis to be a late-time attractor.

The lines $L^\pm$ correspond to the spatially flat, dilaton-vacuum
solutions (\ref{dmv}). The corresponding stable solutions are in 
the range $h_*>\frac{-1}{3}$ for 
$L^+$ and $h_*<\frac{1}{3}$ for $L^-$, respectively. 

\subsubsection{The Invariant Set $\tilde K=0$ for $\Lambda>0$ \label{K0Lp}}

This invariant set was studied in paper I, and the
dynamics there used the variables $\{X_1,Z_1,\nu\equiv
1-X_1^2-Z_1-V_1\}$.  It was found that $Z_1$ is a {\em
monotonically increasing} function (as it is in the full
four-dimensional set).  The early time behaviour of most trajectories
is to asymptote towards the linear dilaton--vacuum solution 
(\ref{static}), represented by the point $C$, 
where all degrees of freedom except
$\varphi$ are dynamically static. To the future, these
solutions asymptote towards the line $L^+$ for $X_1>-\frac{1}{\sqrt{3}}$.  
We note that the point $C$ is the $Y_1=1$ endpoint of the line $L_1$.
Fig.  1 depicts this phase space.
 \begin{figure}[htp]
  \centering
   \includegraphics*[width=4in]{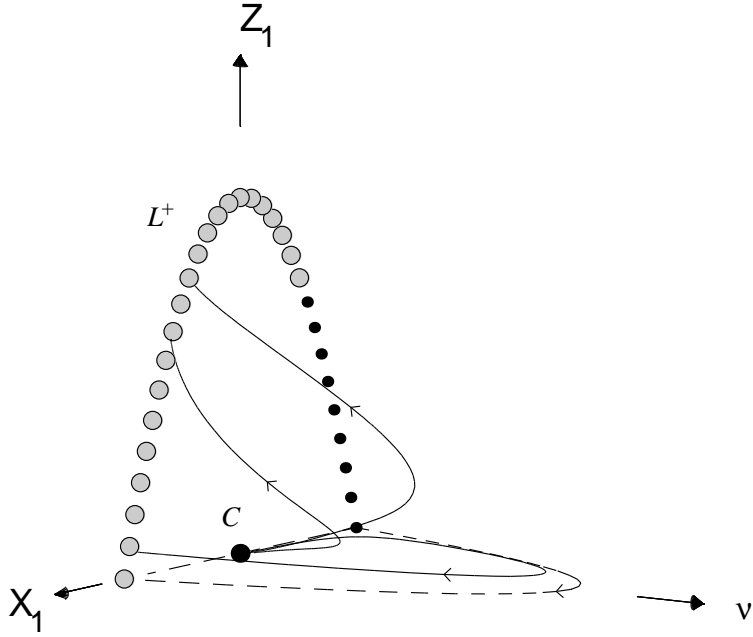}
  \caption{\em Phase diagram of the system
(\ref{dX_1})-(\ref{dV_1}) in the NS-NS ($\Lambda>0$) sector with
$\rho\neq0$ and $\tilde K=0$.  The label $L^+$ refers to a {\it line}
of equilibrium points.  In this phase space, $\dot\varphi>0$ is
assumed.  The labels in all figures correspond to those equilibrium
points discussed in the text.  Throughout, large black dots will
represent sources (i.e., repellers), large grey-filled dots will
represent sinks (i.e., attractors), and small black dots will
represent saddles.  Grey lines represent typical trajectories found
within the two-dimensional invariant sets, dashed black lines are
those trajectories along the intersection of the invariant sets, and
solid black lines are typical trajectories within the full
three-dimensional phase space.}
 \end{figure}

\subsubsection{The Invariant Set $\rho=0$ for $\Lambda>0$, $\tilde K>0$}

In the $\rho=0$ case, the system reduces to the three dimensions of
$\{X_1,Y_1,Z_1\}$ ($V_1=1-X_1^2-Z_1$).  The equilibrium points are the
lines $L^\pm$ with eigenvalues $\lambda_1$, $\lambda_2$ and
$\lambda_3$ (from Section 3.1), and the two endpoints of $L_1$
with eigenvalues $\lambda_\pm$ and $\lambda_3$ (from Section 3.1).
These endpoints are specified by the condition 
$Y_1^2=1$ and we denote them by $L_1^{(\pm)}$,
where the ``$\pm$'' in the superscript reflects the sign of $Y_1$. 

For this invariant set the entire line $L^+$ acts as a
global sink and the entire line $L^-$ acts as a global source.
Furthermore, we note that for $L_1^{\pm}$, $\lambda_-=0$, and so these two
points are non-hyperbolic.  However, the eigenvectors associated with
these zero eigenvalues are both $[\frac{-2}{\sqrt{3}},1,0]$ and are 
completely located in the $(X_1,Y_1)$ plane.  Hence, if we choose $Z_1=0$ and
rotate the $(X_1,Y_1)$ axes such that 
\begin{displaymath}
\tilde x \equiv (Y_1\mp 1)-\frac{\sqrt3}{2}X_1, \qquad 
\tilde y \equiv (Y_1\mp 1)+\frac{2}{\sqrt3}X_1,
\end{displaymath}
we see that in the vicinity of the equilibrium point, the 
trajectories along $\tilde x$ for $\tilde y=0$ are also along
these eigenvectors.  Hence, 
for $\tilde y=0$ and small $\tilde x$, it follows that
\begin{displaymath}
\frac{d\tilde x}{dT} \approx \mp \frac{\tilde x}{7},
\end{displaymath}
Consequently, for $Y_1=+1$, the trajectory along $\tilde x$
asymptotes towards the equilibrium point, whereas the trajectory along
$\tilde x$ for $Y_1=-1$ asymptotes away from the equilibrium point. This 
implies that the points $L_1^{\pm}$ are saddle points.  The phase 
space is depicted in Fig.  2. 

The quantity $X_1/\sqrt{Z_1}$ is monotonically 
decreasing.  Such a monotonic
function excludes the possibility of periodic or recurrent orbits in
this three-dimensional space.  Therefore, solutions generically
asymptote into the past towards $L^-$ 
and into the future towards $L^+$. 
In this three-dimensional set, spatial  curvature is
dynamically important only at intermediate times.
 \begin{figure}[htp]
  \centering
   \includegraphics*[width=5in]{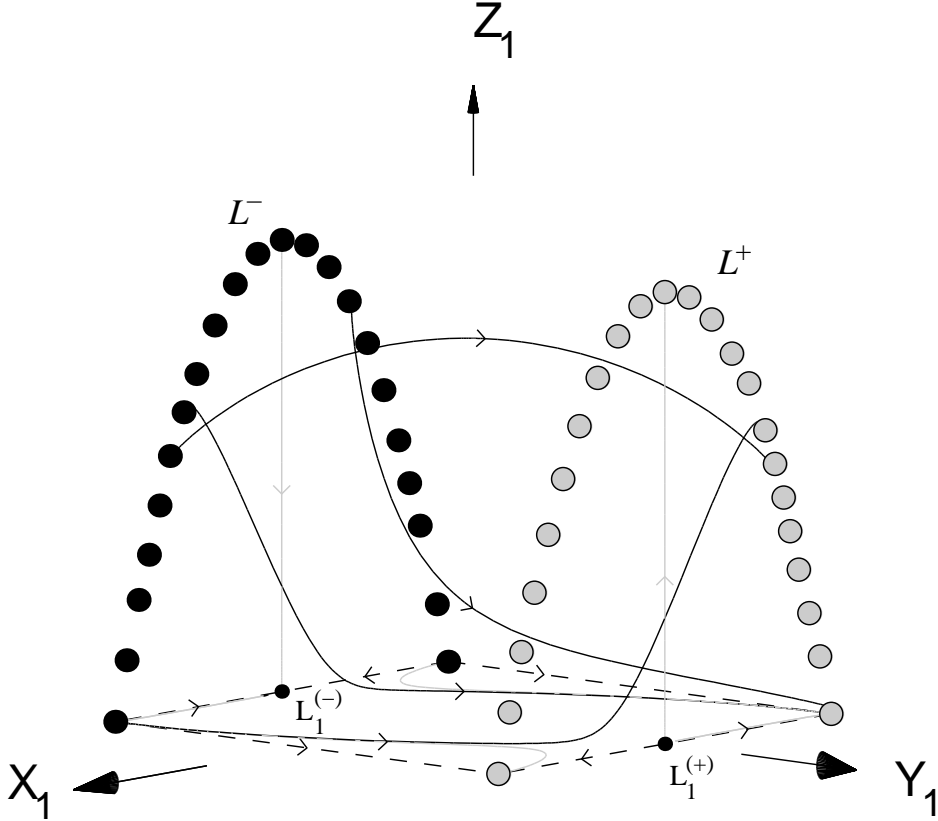}
  \caption{\em Phase diagram of the system
(\ref{dX_1})-(\ref{dV_1}) in the NS-NS ($\Lambda>0$) sector with
$\rho=0$ and $\tilde K>0$. The labels $L^+$ and $L^-$ refer to  {\it
lines} of equilibrium points, and the labels $L_1^{(+)}$ and
$L_1^{(-)}$ represent the equilibrium points which are the endpoints
of the line $L_1$ (for which $X_1=+1$ and $X_1=-1$, respectively).  In
this phase space, $\dot\varphi>0$ is assumed.  See also caption to
Fig.  1.}
 \end{figure}

\subsubsection{Qualitative Analysis of the Four-Dimensional System}

The qualitative dynamics in the full four-dimensional phase space is
as follows. The past attractors
are the line $L^-$ for $X_1<\frac{1}{\sqrt 3}$ and the line $L_1$ for $Y_1>0$.
The future-attractor sets are the lines $L^+$ for $X_1>\frac{-1}{\sqrt 3}$
and the line $L_1$ for $Y_1<0$.  We note that
$Y_1/\sqrt{V_1}$ is monotonically increasing and this implies that 
there are no periodic or recurrent orbits in the full
four--dimensional phase space.  Therefore, solutions are generically
asymptotic in the past to either the line $L^-$ for 
$h_*<\frac{1}{3}$ or to the line $L_1$ for $n>0$.  Similarly,
solutions are generically asymptotic in the future to either 
$L^+$ for $h_*>\frac{-1}{3}$ or $L_1$ for $n<0$.  

Since $Y_1/\sqrt{V_1}$ is monotonically increasing, $Y_1\rightarrow
+1$ or $V_1\rightarrow 0$ asymptotically to the future. These limits 
represent global sinks on the line $L^+$. Conversely, 
$Y_1\rightarrow -1$ or
$V_1\rightarrow 0$ asymptotically to the past, corresponding to the
global sources on the line $L^-$.  
There are also equilibrium points
for finite $Y_1\equiv Y_*$ inside the phase space on the line $L_1$.
Again, since $Y_1/\sqrt{V_1}$ is monotonically increasing, the points
$Y_*<0$ are global sinks and $Y_*>0$ are global sources.  All of this
is consistent with the above discussion presented in Section 2.2 
regarding the generic asymptotic behaviour.

We note that the reflections $X_1\rightarrow -X_1$ and $Y_1\rightarrow
-Y_1$ are equivalent to a time reversal of the dynamics.  Therefore,
there are orbits starting on the line $L^-$ (for $X_1<\frac{1}{\sqrt 3}$) and
ending on the line $L_1$ (for $Y_*<0$).  Similarly, there are orbits
which begin on the line $L_1$ (for $Y_*>0$) and end on the line $L^+$
(for $X_1>\frac{-1}{\sqrt 3}$).
Due to the existence of the monotonic function and the
continuity of orbits in the four-dimensional phase space, solutions
cannot start and finish on $L_1$.  This is best illustrated in the
invariant set $Z_1=0$.  In addition, orbits may start on the line $L^-$
(for $X_1<\frac{1}{\sqrt 3}$) and end on the line $L^+$ (for $X_1>\frac{-1}{\sqrt
3}$).  Investigation of the invariant set $Z_1=0$ also
indicates which sources and sinks are connected; not all orbits from
$L^-$ can evolve towards $L^+$.

Although the lines $L^\pm$ lie in both of the invariant sets $\rho=0$ and
$\tilde K=0$, the line $L_1$ does not. On this line,  
$X_1=0$, and the 
solutions are therefore static $(\dot\alpha = \dot{\beta} =0)$. 
Eq. (\ref{dchi}) then implies
that the axion field and spatial curvature 
can both be dynamically significant at early 
and late times for the appropriate orbits. 

\subsection{The Case $\Lambda>0$, $\tilde K<0$}

In this case, we choose the negative sign for $U_2$
in Eq. (\ref{TheDefs}), the positive sign for $V_2$, and 
the definition $\xi^2 = \dot\varphi^2$.  The generalized
Friedmann constraint (\ref{rrfriedmann}) now implies that 
\begin{equation}
0 \leq X_2^2 + Z_2 + U_2 + V_2\leq 1.
\end{equation}
For this system, $Y_2^2=1$, and so the
four-dimensional system consists of the variables $0\leq \{X_2^2, Z_2, U_2, 
V_2\} \leq 1$:
\begin{eqnarray}
\label{dX_2}
\frac{dX_2}{dT} & = & \sqrt 3 \left( 1-X_2^2-Z_2-V_2-\frac{2}{3}U_2  \right) 
		     + X_2 \left( 1-X_2^2-Z_2 \right), \\
\label{dZ_2}
\frac{dZ_2}{dT} &=& 2Z_2 \left( 1-X_2^2-Z_2 \right) > 0, \\
\label{dU_2}
\frac{dU_2}{dT} &=& -2U_2 \left(X_2^2+Z_2+\frac{1}{\sqrt3}X_2 \right),\\
\label{dV_2}
\frac{dV_2}{dT} &=& -2V_2 \left( X_2^2 + Z_2\right) < 0.
\end{eqnarray}
The invariant sets $X_2^2+Z_2+U_2+V_2=1$, $Z_2=0$, $V_2=0$, $U_2=0$ define the
boundary of the phase space.  The equilibrium sets and their
corresponding eigenvalues (denoted by $\lambda$) are
\begin{eqnarray}
S^+: & & X_2=-\frac{1}{\sqrt3}, Z_2=0, U_2=\frac{2}{3}, V_2=0; \nonumber \\ &&
   \left(\lambda_1, \lambda_2, \lambda_3, \lambda_4\right) =  \left(
   -\frac{2}{3}, \frac{2}{3}, \frac{4}{3}, \frac{4}{3} \right), \\
C: & & X_2=0,Z_2=0, U_2=0, V_2=1; \nonumber \\&&
   \left(\lambda_1,\lambda_2, \lambda_3, \lambda_4\right) =  \left(
   1,0, 2, 0 \right), \\
L^+: & & Z_2=1-X_2^2, U_2=0, V_2=0; \nonumber \\ &&
   \left(\lambda_1, \lambda_2, \lambda_3, \lambda_4\right) =  \left(
    -\frac{2}{\sqrt3} \left[ X_2+\sqrt3 \right],-2,0,-2\sqrt3 \left[ X_2+\frac{1}{\sqrt3} \right]    \right).
\end{eqnarray}

The point $C$ represents the static `linear dilaton--vacuum' solution
(\ref{static}).  The saddle $S^+$ represents the ``$-$'' branch of the 
Milne solution (\ref{curv_drive}), 
where only the curvature term
and scale factor are dynamic.

\subsubsection{The Invariant Set $\rho=0$ for $\Lambda>0$, $\tilde K<0$}

In the $\rho=0$ case, the system reduces to the three dimensions of
$\{X_2,Z_2,U_2\}$ ($V_2=1-X_2^2-Z_2-U_2$).  The equilibrium points are
the same as above with eigenvalues $\lambda_1$, $\lambda_2$ and
$\lambda_3$.  We note that for this invariant set the entire
line $L^+$ acts as a global sink, and $C$ acts as a source.  Although
one of the eigenvalues for the point $C$ is zero, it is shown in the 
following subsubsection that this point is a 
source in the full four--dimensional phase space. The argument 
is identical in this subsection. 

The variable $Z_2$ is {\em
monotonically increasing}, and as such we see that the shear term is 
negligible at early times, but becomes dynamically significant at late times.
There are no  periodic or recurrent orbits in this three--dimensional
phase space due to the existence of this monotonic function. 
In general, solutions asymptote into the past towards
the static solution (\ref{static}) (point $C$). 
Solutions asymptote into the future 
towards $L^+$.  Fig.  3 depicts this three-dimensional phase space.
 \begin{figure}[htp]
  \centering
   \includegraphics*[width=5in]{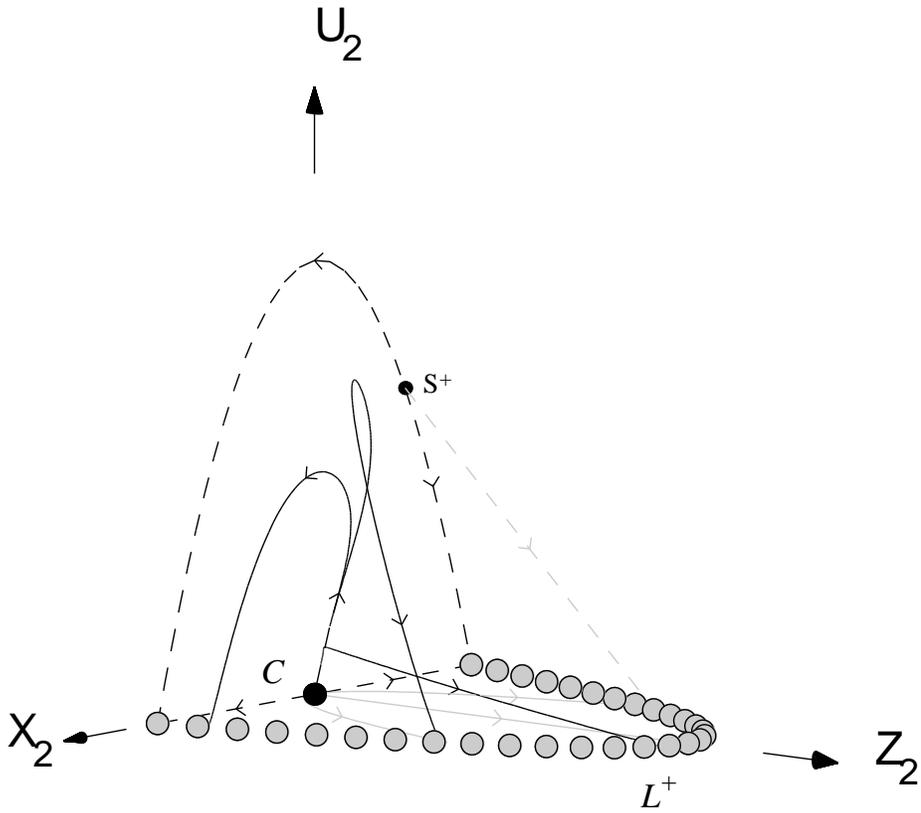}
  \caption{\em Phase diagram of the system
(\ref{dX_2})-(\ref{dV_2}) in the NS-NS ($\Lambda>0$) sector with
$\rho=0$ and $\tilde K<0$. Note that $L^+$ represents a {\em line} of
equilibrium points.  See also caption to Fig.  1.}
 \end{figure}

\subsubsection{Qualitative Analysis of the Four-Dimensional System}

The qualitative behaviour for the invariant set $\tilde K=0$ is
discussed in Section 3.1.1 (see Fig. 1).
In the four-dimensional set, the point $C$ is non-hyperbolic because
of the two zero eigenvalues.  However, it can be shown that this point 
is a source in the four-dimensional set by the following
argument.  We first note that the variable $Z_2$ is {monotonically
increasing} and hence orbits asymptote into the past towards the
invariant set $Z_2=0$.  Similarly, $V_2$ is a {monotonically decreasing}
function, and so orbits asymptote into the past towards large
$V_2$ (i.e., $V_2=1$). This implies that they asymptote towards the point
$C$.

Since $Z_2=0$ asymptotically, let us consider the invariant set $Z_2=0$,
where Eqs. (\ref{dX_2})-(\ref{dV_2}) become:
\begin{eqnarray}
\label{dX}
\frac{dX_2}{dT} & = & \sqrt 3 \left( 1-X_2^2-V_2-\frac{2}{3}\tilde U_2 \right)
	+ X_2 \left( 1-X_2^2\right), \\
\label{dU}
\frac{dU_2}{dT} &=& -2U_2 \left(X_2^2+\frac{1}{\sqrt3}X_2 \right),\\
\label{dV}
\frac{dV_2}{dT} &=& -2 V_2 X_2^2.
\end{eqnarray}
It is clear from Eq. (\ref{dV}) that $V_2$ increases monotonically into
the past.  Now, this
three-dimensional phase space is bounded by the surface
$X_2^2+U_2+V_2=1$, the ``apex'' of which lies at $V_2=1$ (and
$X_2=U_2=0$).  Therefore, all orbits in or on this phase space boundary
lie below $V_2=1$, and therefore asymptote into the
past towards $V_2=1$.
To further illustrate that this point is indeed a source, it is helpful 
to consider the
invariant set $Z_2=U_2=0$, $X_2^2+V_2=1$.  In the neighbourhood of $C$, 
Eq. (\ref{dX}) becomes $dX_2/dT = X_2(1-X_2^2)$,
indicating that orbits are repelled from $X_2=0$.  Hence, the point
$C$ is the past attractor to the full four-dimensional set.  

The future attractor for this set is the line $L^+$ (for
$X_2>\frac{-1}{\sqrt{3}}$).  Both $C$ and $L^+$ lie in both of the
invariant sets $\rho=0$ and $\tilde K=0$, which is consistent with the
analysis of Eq. (\ref{dchi}).  We conclude, 
therefore, that the spatial 
curvature terms and the axion field are dynamically important only at
intermediate times, and are negligible at early and late
times.  The dynamical effect of the shear becomes increasingly important
because the variable $Z_2$ increases monotonically. On the other hand, 
the variable $V_2$ decreases monotonically and the dynamical
effect of the central charge deficit, $\Lambda$, becomes increasingly
negligible.  In addition, the existence of monotone
functions in the four--dimensional phase space prohibits closed orbits and
serves as proof of the evolution that is described above.

\subsection{The Case $\Lambda<0$, $\tilde K>0$}

We choose the positive sign for $U_3$ and the negative
sign for $V_3$ in Eq. (\ref{TheDefs}) to ensure that
these variables are positive definite. We also define 
$\xi^2 \equiv  \dot\varphi^2+3\tilde K-2\Lambda$ for this case. The 
generalized Friedmann constraint equation can then be rewritten as
\begin{equation}
0 \leq X_3^2 + Z_3 \leq 1,\qquad Y_3^2+U_3+V_3=1,
\end{equation}
We again eliminate $U_3$ (which is proportional to $\tilde K$), and 
consider the
four-dimensional system of ODEs for $0\leq \{X_3^2, Y_3^2, Z_3, V_3\} \leq1$:
\begin{eqnarray}
\label{dX_3}
\frac{dX_3}{dT} & = & \left(1-X_3^2-Z_3\right)\!\! \left(\sqrt 3 +X_3Y_3\right)
-\!\frac{1}{\sqrt 3}\left(1-Y_3^2-V_3 \right)\!\! \left(1-X_3^2-Z_3\right), \\
\label{dY_3}
\frac{dY_3}{dT} &=& \frac{1}{\sqrt3}X_3Y_3 \left( 1-Y_3^2 -V_3 
	\right) +\left(1-Y_3^2 \right)\left(X_3^2+Z_3\right) ,\\
\label{dZ_3}
\frac{dZ_3}{dT} &=& 2Z_3\left[ \frac{1}{\sqrt3}X_3 
	\left( 1-Y_3^2-V_3 \right) +Y_3\left( 1-X_3^2-Z_3 
	\right) \right], \\
\label{dV_3}
\frac{dV_3}{dT} &=& 2V_3 \left[ \frac{1}{\sqrt3} X_3
	\left(1-Y_3^2-V_3 \right) - Y_3\left( X_3^2 + Z_3
	\right) \right].
\end{eqnarray}
The invariant sets $Y_3^2+V_3=1$, $X_3^2+Z_3=1$, $Z_3=0$ define the
boundary of the phase space.  The equilibrium sets and their
corresponding eigenvalues (denoted by $\lambda$) are
\begin{eqnarray}
L^\pm: & & Y_3=\pm 1, Z_3=1-X_3^2, V_3=0; \nonumber \\ &&
   \left(\lambda_1, \lambda_2, \lambda_3, \lambda_4\right) =  \left(
   \mp\frac{2}{\sqrt3} \left[ \sqrt3 \pm X_3\right], \mp2, 0, \mp2\sqrt3
	\left[ \frac{1}   {\sqrt3} \pm X_3 \right]    \right), 
\end{eqnarray}
where again the zero eigenvalues arise because these are all {\em
lines} of equilibrium points.  Here, the global sink is the line
$L^+$ for $X_3>\frac{-1}{\sqrt{3}}$ (saddle otherwise), and the global source
is the line $L^-$ for $X_3< \frac{1}{\sqrt{3}}$ (saddle otherwise). Stable 
solutions on the line 
$L^+$ correspond to the range 
$h_*>\frac{-1}{3}$. The stable solutions on $L^-$
arise when $h_*<\frac{1}{3}$.

\subsubsection{The Invariant Set $\tilde K=0$\label{K0Ln} for $\Lambda<0$}

This invariant set was studied in paper I, the
dynamics of which are as follows. The
variables $X_3$ and $Y_3$ are {monotonically increasing} functions,
corresponding to $\dot\alpha$ and $\dot\varphi$ (respectively). The 
former implies that these trajectories represent 
cosmologies that are initially 
contracting and then reexpand.  In paper I,
the third variable used was $\kappa\equiv 1-X_3^2-Z_3$. This is  
proportional to $\rho$ and is only 
dynamically significant at intermediate times. It asymptoted to zero
into the past and future, indicating that the axion field is
negligible at early and late times.  All the equilibrium points in this
invariant set are represented by the dilaton-vacuum solutions
(\ref{dmv}) (the lines $L^{\pm}$). In general, orbits asymptote
into the past towards the line $L^-$ (for $X_3< \frac{1}{\sqrt{3}}$),
and asymptote to the future towards the line $L^+$ (for
$X_3>\frac{-1}{\sqrt{3}}$).  Fig.  4 depicts this three-dimensional
phase space, using the variables $\{X_3,Y_3,\kappa\}$.
 \begin{figure}[htp]
  \centering
   \includegraphics*[width=4in]{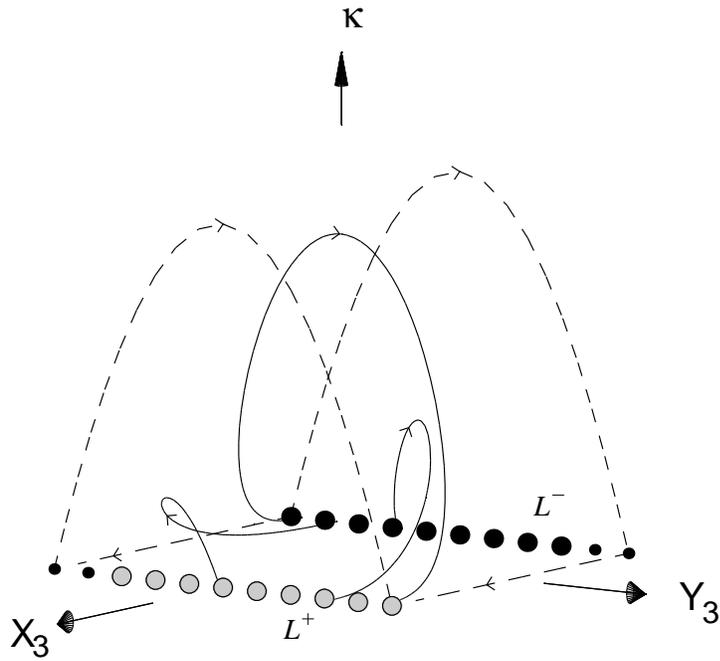}
  \caption{\em Phase portrait of the system
(\ref{dX_3})-(\ref{dV_3}) in the NS-NS ($\Lambda<0$) sector for $K=0$
and $\rho\neq 0$.  Note that the labels $L^+$ and $L^-$ refer to {\em lines} of
equilibrium points.  See also caption to Fig.  1.}
 \end{figure}

\subsubsection{The Invariant Set $\rho=0$ for $\Lambda<0$, $\tilde K>0$}

Since $\dot\alpha\neq0$ at the equilibrium points, we examine the
$\rho=0$ case, where the system reduces to the three dimensions
$\{X_3,Y_3,V_3\}$ ($Z_3=1-X_3^2$).  The equilibrium points are the
same as above with eigenvalues $\lambda_1$, $\lambda_2$, $\lambda_3$.
We note that the
entire line $L^+$ acts as a global sink and that the entire line $L^-$
acts as a global source in this invariant set.  The
function $Y_3/\sqrt{V_3}$ is {monotonically
increasing}, eliminating the possibility of
periodic orbits. 
Fig.  5 depicts this three-dimensional phase space.  
 \begin{figure}[htp]
  \centering
   \includegraphics*[width=5in]{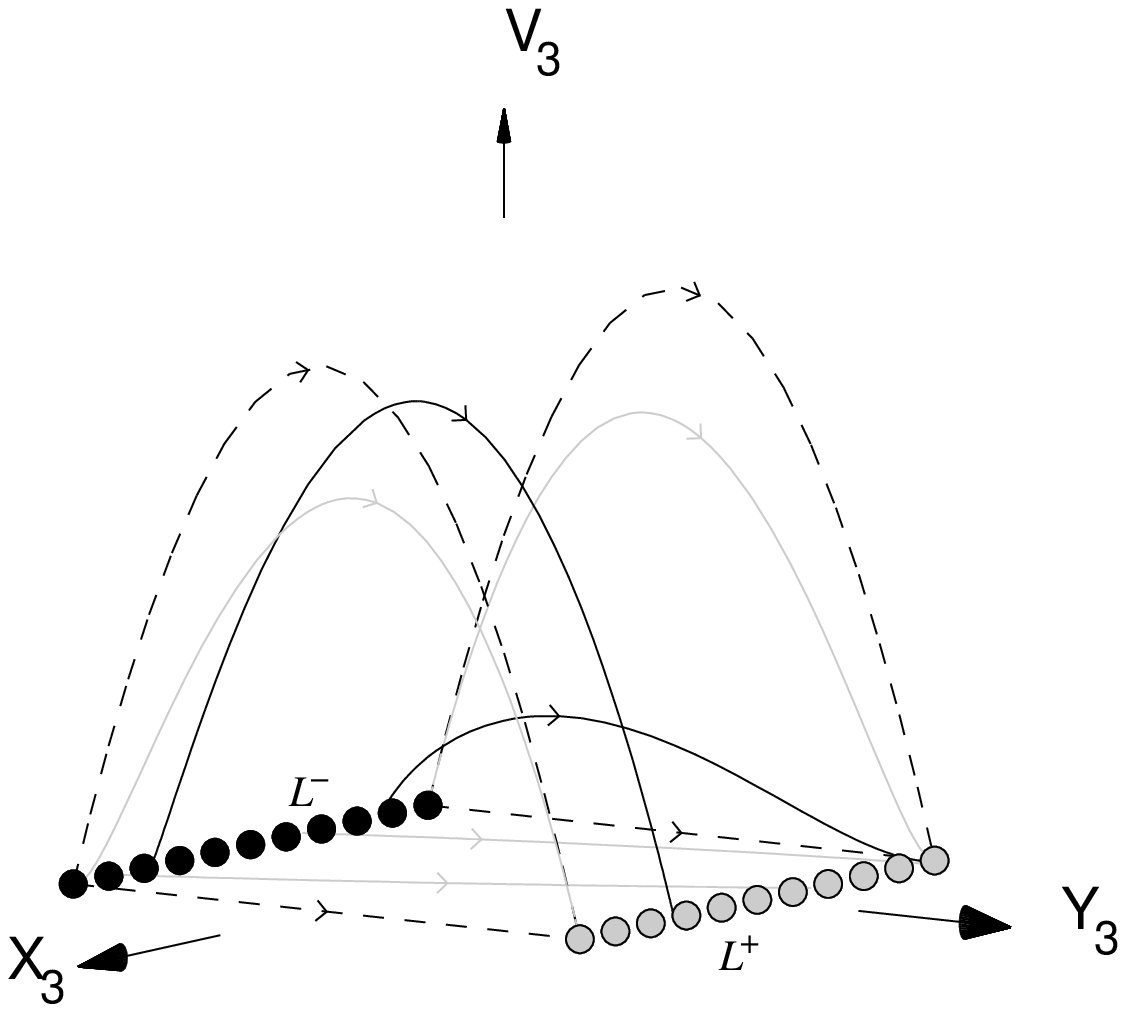}
  \caption{\em Phase diagram of the system
(\ref{dX_3})-(\ref{dV_3}) in the NS-NS ($\Lambda<0$) sector for
$\rho=0$ and $\tilde K>0$.  Note that $L^+$ and $L^-$  represent {\em lines} of
equilibrium points.  See also caption to Fig.  1.}
 \end{figure}

\subsubsection{Qualitative Analysis of the Four-Dimensional System}

The qualitative dynamics in the full four-dimensional phase space is
as follows.  The global repellers and attractors 
are the lines $L^-$ and $L^+$, respectively. Orbits generically asymptote
into the past towards the line $L^-$ (for $X_3<\frac{1}{\sqrt 3}$), and
into the future towards $L^+$ (for
$X_3>\frac{-1}{\sqrt 3}$). In the former case the stable solutions 
are given by the ``$+$' branch of Eq. (\ref{dmv}) for $h_*<\frac{1}{3}$. The 
stable sinks are given by the ``$-$'' branch with $h_*>\frac{-1}{3}$. 
Again, we see
that the curvature term, axion field and the central charge
deficit are dynamically important only at intermediate times.
The existence of the {\em
monotonically increasing} function
$Y_3/\sqrt{V_3}$ excludes the possibility of
periodic orbits and serves to verify the above description of the 
evolution of the solutions in the four--dimensional set.

\subsection{The Case $\Lambda<0$, $\tilde K<0$}

For this case, the appropriate definition 
for the variable $\xi$ is $\xi^2 = \dot\varphi^2-2\Lambda$ and 
we choose the negative signs for
both $U_4$ and $V_4$ in Eq. (\ref{TheDefs}). The generalized Friedmann
constraint equation is written as 
\begin{equation}
0 \leq X_4^2 + Z_4 + U_4\leq 1, \qquad Y_4^2+V_4=1,
\end{equation}
Treating $V_4$ as the extraneous variable results in 
the four-dimensional system consisting of the variables $0\leq \{X_4^2, 
Y_4^2, Z_4, U_4\} \leq 1$:
\begin{eqnarray}
\label{dX_4}
\frac{dX_4}{dT} & = & \left(1-X_4^2-Z_4\right)\left(\sqrt 3 +X_4Y_4\right) 
-\frac{2}{\sqrt 3}U_4, \\
\label{dY_4}
\frac{dY_4}{dT} &=& \left(1-Y_4^2\right) \left( X_4^2 + Z_4\right) > 0, \\
\label{dZ_4}
\frac{dZ_4}{dT} &=& 2Z_4Y_4 \left( 1-X_4^2-Z_4 \right), \\
\label{dU_4}
\frac{dU_4}{dT} &=& -2U_4 \left[ Y_4\left(X_4^2+Z_4\right)
	+\frac{1}{\sqrt3}X_4 \right].
\end{eqnarray}

The invariant sets $X_4^2+Z_4+U_4=1$, $Z_4=0$, $Y_4^2=1$, $U_4=0$ define the
boundary of the phase space.  The equilibrium sets and their
corresponding eigenvalues (denoted by $\lambda$) are
\begin{eqnarray}
S^\pm: & & X_4=\mp\frac{1}{\sqrt3}, Y_4=\pm1, Z_4=0, U_4=\frac{2}{3}; \nonumber \\ &&
   \left(\lambda_1, \lambda_2, \lambda_3, \lambda_4\right) =  \left(
   \pm\frac{2}{3}, \mp\frac{2}{3}, \pm\frac{4}{3}, \pm\frac{4}{3} \right), \\
L^\pm: & & Y_4=\pm1, Z_4=1-X_4^2, U_4=0; \nonumber \\ &&
   \left(\lambda_1, \lambda_2, \lambda_3, \lambda_4\right) =  \left(
   \mp\frac{2}{\sqrt3} \left[ \sqrt3 \pm X_4\right], \mp 2 , 0,
  \mp 2\sqrt3 \left[ \frac{1}{\sqrt3} \pm X_4 \right] \right).
\end{eqnarray}
The global source for this system is the line $L^-$ (for
$X_4<\frac{1}{\sqrt{3}}$, $h_*<\frac{1}{3}$)
and the global sink is the line $L^+$ (for
$X_4>\frac{-1}{\sqrt{3}}$, $h_*>\frac{-1}{3}$). 
The saddle points, 
$S^\pm$, are represented by the Milne models 
(\ref{curv_drive}), where $S^+$ corresponds to the
``$-$'' solution and $S^-$ to the ``$+$'' solution.

\subsubsection{The Invariant Set $\rho=0$ for $\Lambda <0$, $\tilde K<0$}

This system reduces to the three dimensions of
$\{X_4,Y_4,Z_4\}$ ($U_4=1-X_4^2-Z_4$).  The equilibrium points are the
same as above with eigenvalues $\lambda_1$, $\lambda_2$, $\lambda_3$.
The entire lines $L^+$ and $L^-$ now act
as a global sink and source, respectively. Recurrent orbits
are forbidden by the existence of the monotonically 
increasing variable $Y_4$. Hence, solutions
generically asymptote into the past (future) towards the ``$+$''
(``$-$'') dilaton--vacuum solutions (\ref{dmv}) and the curvature
term and central charge deficit are dynamically significant only at
intermediate times.  Fig.  6 depicts this three-dimensional phase
space.
 \begin{figure}[htp]
  \centering
   \includegraphics*[width=5in]{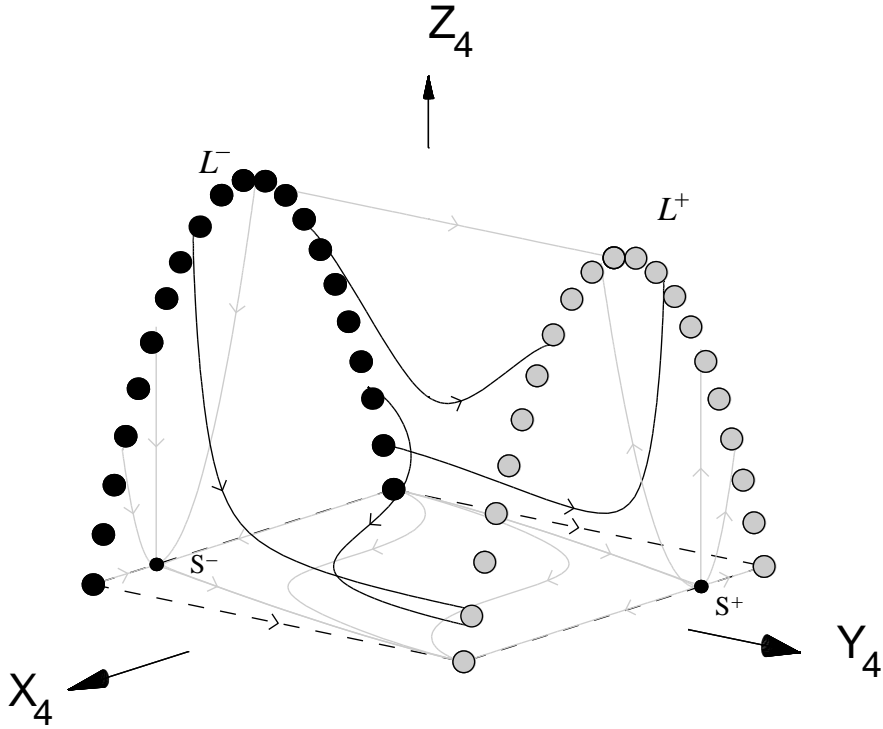}
  \caption{\em Phase portrait of the system
(\ref{dX_4})-(\ref{dU_4}) in the NS-NS ($\Lambda<0$) sector with $\rho= 0$ and
$\tilde K<0$.  Note that $L^+$ and $L^-$  represent {\em lines} of
equilibrium points.  See also caption to Fig.  1.}
 \end{figure}

\subsubsection{Qualitative Analysis of the Four-Dimensional System}

The dynamical behaviour in the invariant set $\tilde K=0$ is identical
to that described in Section 3.3.1 (see Fig. 4).
The qualitative dynamics in the full four-dimensional phase space is
as follows.  Since $Y_4$ is monotonically increasing, 
the orbits asymptote into the past towards $Y_4=-1$ and into the future
towards $Y_4=+1$.  
As in the above examples, 
the existence of such a monotonic function
excludes the possibility of periodic orbits in the four--dimensional
phase space.  Most orbits asymptote into the past towards the line
$L^-$ (for $X_4<\frac{1}{\sqrt{3}}$), and into the future towards the line $L^+$
(for $X_4>\frac{-1}{\sqrt{3}}$). 
The range of values for $h_*$ corresponding 
to stable solutions is given by $h_* > \frac{-1}{3}$ for those on $L^+$ 
and $h_*<\frac{1}{3}$ for those on $L^-$.  The variable $Y_4$ increases
monotonically along orbits and this implies that 
$\dot\varphi$ is dynamically significant
at early and late times, but dynamically insignificant at intermediate
times.  Conversely, we see that the curvature term, central charge
deficit and axion field are dynamically important only at
intermediate times, and are negligible at early and late times.

This concludes the qualitative analysis of the isotropic curvature 
string cosmologies with NS--NS fields. In the next Section, we 
consider the effect on the dynamics of introducing a non--trivial 
$\Lambda_{\rm M}$.

\setcounter{equation}{0}
\section{Non-Zero Cosmological Constant $\Lambda_{\rm M}$ ($\Lambda=0$)}
 
We begin this Section by defining new variables
\begin{equation}
\label{beta}
\frac{d}{dt}\equiv e^{\frac{1}{2}(\varphi+3\alpha)}\frac{d}{dT}, \quad
N \equiv 6{\beta'}^2 , 
\qquad \psi \equiv \varphi' , \qquad h \equiv \alpha', \qquad K \equiv
\tilde K e^{-(\varphi+3\alpha)},
\end{equation}
where a prime denotes differentiation with respect to the new time
coordinate, $T$.
Eqs. (\ref{rr1})--(\ref{rrfriedmann}) may then 
be written as the set of ODEs: 
\begin{eqnarray}
\label{rr1a}
h ' &=& -\frac{3}{2} h^2+\frac{1}{2}h\psi - K -\frac{1}{2} \Lambda_{\rm M} +
		\frac{1}{2}\rho e^{-(\varphi+3\alpha)} \\
\label{rr2a}
\psi' &=& \frac{3}{2}h^2 -\frac{3}{2}h\psi +\frac{1}{2} N+ \frac{3}{2} K
		- \frac{1}{4} \rho e^{-(\varphi+3\alpha)} \\
\label{rr3a}
N' &=& (\psi-3h) N \\
\label{rr4a}
K' &=& -(\psi + 5h) K \\ 
\label{rr5a}
\rho' &=& -6h \rho \\
\label{rrfriedmanna}
3h^2 &-&\psi^2 +N -3K +\Lambda_{\rm M} +\frac{1}{2}\rho e^{-(\varphi+3\alpha)}=0 
\end{eqnarray}

The variable $\rho$ may be eliminated due to 
the constraint equation (\ref{rrfriedmann}). It also proves 
convenient to further define a set of variables 
\begin{equation}
\mu\equiv \frac{\sqrt3h}{\xi}, \quad
\chi\equiv \frac{\psi}{\xi}, \quad
\nu\equiv \frac{N}{\xi^2}, \quad
\zeta\equiv \frac{\pm 3 K}{\xi^2}, \quad
\lambda\equiv \frac{\pm \Lambda_{\rm M}}{\xi^2}, \quad
\frac{d}{dT}\equiv \xi \frac{d}{d\tau},\label{TheDefs2}
\end{equation}
where the $\pm$ signs ensure that 
$\zeta>0$ and $\lambda>0$.  
The variable $\xi$ is defined in each of the following
four subsections by 
\begin{itemize}
\item $\Lambda_{\rm M}>0$\begin{itemize} \item $K>0$: $\xi^2 \equiv 3K
	+\psi^2$  (Section 4.1),
\item $K<0$: $\xi^2 \equiv \psi^2$  (Section 4.2),\end{itemize}
\item $\Lambda_{\rm M}<0$ \begin{itemize} \item $ K>0$: $\xi^2 \equiv 3 K
	+\psi^2-\Lambda_{\rm M}$  (Section 4.3),
\item $K<0$: $\xi^2 \equiv \psi^2-\Lambda_{\rm M}$ (Section 4.4).
\end{itemize}\end{itemize}
With these
definitions, all variables are bounded, 
$0\leq \{\mu^2, \chi^2, \nu, \zeta,
\lambda\} \leq 1$, and Eq. (\ref{rrfriedmanna}) now reads
\begin{equation}
\frac{1}{2}\rho\xi^{-2} e^{-(\varphi+3\alpha)} = \chi^2 \pm \zeta \mp \lambda 
  -\mu^2-\nu >0.\label{newFried2}
\end{equation}
Our overall approach in this Section 
is identical to that of Section 3 and we refer the reader to 
the discussion immediately after Eq. (\ref{cf}) for the 
general outline  adopted.

\subsection{The Case $\Lambda_{\rm M}>0$, $K>0$}

For $K>0$, Eq. (\ref{rrfriedmanna}) is written in the new variables as
\begin{equation}
0\leq \mu_1^2+\nu_1+\lambda_1\leq 1, \qquad \zeta_1+\chi_1^2=1,
\end{equation}
where the ``$+$'' sign is chosen for both $\lambda$ and $\zeta$ in Eq. 
(\ref{TheDefs2}).  For this case, the variable $\zeta_1\equiv 1-\chi_1^2$ will
be considered extraneous and the system (\ref{rr1a})-(\ref{rr4a}) then
reduces to the four-dimensional system:
\begin{eqnarray}
\label{dmu_1}
\frac{d\mu_1}{d\tau} &=& \left(1-\!\mu_1^2-\!\nu_1-\!\frac{1}{2}\lambda_1
\right)\!
\left(\sqrt 3+\!\mu_1\chi_1\right)-\frac{1}{\sqrt3}\left(1-\!\mu_1^2\right)\!
\left(1-\!\chi_1^2\right)-\sqrt3\lambda_1, \\
\label{dchi_1}
\frac{d\chi_1}{d\tau} & = & \left(1-\chi_1^2\right)\left[\mu_1^2+\nu_1
	+\frac{1}{2}\lambda_1+\frac{1}{\sqrt 3}\mu_1\chi_1\right], \\
\label{dnu_1}
\frac{d\nu_1}{d\tau} &=& \nu_1\left[\frac{2}{\sqrt 3} \mu_1\left(1
	-\chi_1^2\right)+2\chi_1\left(1-\mu_1^2-\nu_1
	-\frac{1}{2}\lambda_1\right)\right], \\ 
\label{dlambda_1}
\frac{d\lambda_1}{d\tau} &=& \lambda_1\left[\frac{1}{\sqrt 3}\mu_1
	\left(5-2\chi_1^2\right) +\chi_1\left(1-2\mu_1^2-2\nu_1
	-\lambda_1\right)\right].
\end{eqnarray}
The invariant sets $\mu_1^2+\nu_1+\lambda_1=1$ ($\rho=0$),
$\chi_1^2=1$ ($K=0$), $\nu_1=0$ ($N=0$) and $\lambda_1=0$
($\Lambda_{\rm M}=0$) define the boundaries to the phase space.  
The equilibrium points and their respective eigenvalues
(denoted by $\lambda$) are given by
\begin{eqnarray}
L^\pm: && \chi_1=\pm1, \mu_1^2+\nu_1=1, \lambda_1=0; \nonumber \\ \nonumber &&
  (\lambda_1,\lambda_2,\lambda_3,\lambda_4) = \left (0,
   \mp\frac{2}{\sqrt 3}\left[\sqrt 3\pm \mu_1\right], \sqrt 3 \left[\mu_1\mp 
     \frac{1}{\sqrt 3}\right],\mp2\sqrt{3}\left[\frac{1}{\sqrt 3}\pm\mu\right]\right).\\ \label{Lpm}\\
\nonumber
S^\pm_1: & & \mu_1=\mp\frac{1}{\sqrt{27}}, \chi_1=\pm1, \nu_1=0, \lambda_1=\frac{16}{27}; \nonumber \\ &&
   \left(\lambda_1,\lambda_2, \lambda_3, \lambda_4\right) =  \left(
   \mp\third\left[1+i\frac{\sqrt{231}}{3}\right],\mp\third\left[1-i\frac{\sqrt{231}}{3}\right], \mp\frac{4}{3}, \pm\frac{4}{9} \right).
\end{eqnarray}
{}From the eigenvalues, we deduce that $L^+$ is a late-time attractor
for $\mu_1^2< \frac{1}{3}$, and that $L^-$ is an early-time repeller for
$\mu_1^2<\frac{1}{3}$.  In both cases, $\chi_1^2=1$ and therefore $K=0$
($\zeta_1=0$).  The ``$+$'' solution of (\ref{dmv}) is a sink for
$h_*^2<\frac{1}{9}$ and the ``$-$'' solution of (\ref{dmv}) is a
source for $h_*<\frac{1}{9}$.  The points $S^\pm_1$ are saddle points
on the boundary of the phase space and correspond to the exact
solution (\ref{newsol}).

\subsubsection{The Invariant Set $K=0$ for $\Lambda_{\rm M}>0$ \label{K0LRp}}

This invariant set was studied in paper II by employing 
dynamical variables equivalent to $\{\mu_1,\lambda_1,\nu_1\}$, and we
now summarize the important features of the model. 
For spatially isotropic solutions confined to the invariant set
$\nu_1=0$, most trajectories evolve from the equilibrium point
$S^+_1$ (labelled ``$F$'' in paper II) located at
$(\mu_1,\nu_1,\lambda_1)=(\frac{-1}{\sqrt{27}},0,\frac{16}{27})$ (corresponding to the
``$-$'' solution (\ref{newsol})) and are future
asymptotic to a heteroclinic orbit. There are two saddle 
equilibrium points, $S_1$ and $S_2$, which are the
endpoints of the line $L^+$; $S_1$ is given by the ``$-$'' 
branch of Eq. (\ref{dmv}) with 
$h_*=\frac{-1}{\sqrt{3}}$ and $S_2$ corresponds to 
$h_*=\frac{1}{\sqrt{3}}$. There are also the 
single boundary orbits in the invariant sets
$\lambda_1=0$, corresponding to $\Lambda_{\rm M}=0$  and 
$\lambda_1+\mu_1^2=1$ (constant axion field).
An orbit spends the majority of its time in the neighbourhoods 
of $S_1$ and $S_2$ 
and shadows the respective boundary orbits as it rapidly moves between 
the two saddles. Progressively more time is spent 
near the saddles for each completed cycle, and the dynamics 
is therefore not periodic. 

An anisotropic contribution $(\nu_1 \ne 0)$ does not introduce 
new sources into the system and the point $S_1^+$ is still 
the only source.  Fig.  7 depicts the full three-dimensional space. 
Eq. (\ref{dnu_1}) implies that $\nu_1$ is a monotonically 
increasing function and consequently the orbits are 
repelled from $S_1^+$ and spiral out 
monotonically. 
The general behaviour for most trajectories in this phase space is to
evolve away from the equilibrium point $S_1^+$, spiral about the
line $\mu_1=\frac{-1}{\sqrt{27}}$, 
$\nu_1=\frac{13}{8}(\lambda_1-\frac{16}{27})$ and eventually 
asymptote towards the line $L^+$ for $\mu_1^2< \frac{1}{3}$.
These trajectories were discussed in detail 
in paper II. 
 \begin{figure}[htp]
  \centering
   \includegraphics*[width=4in]{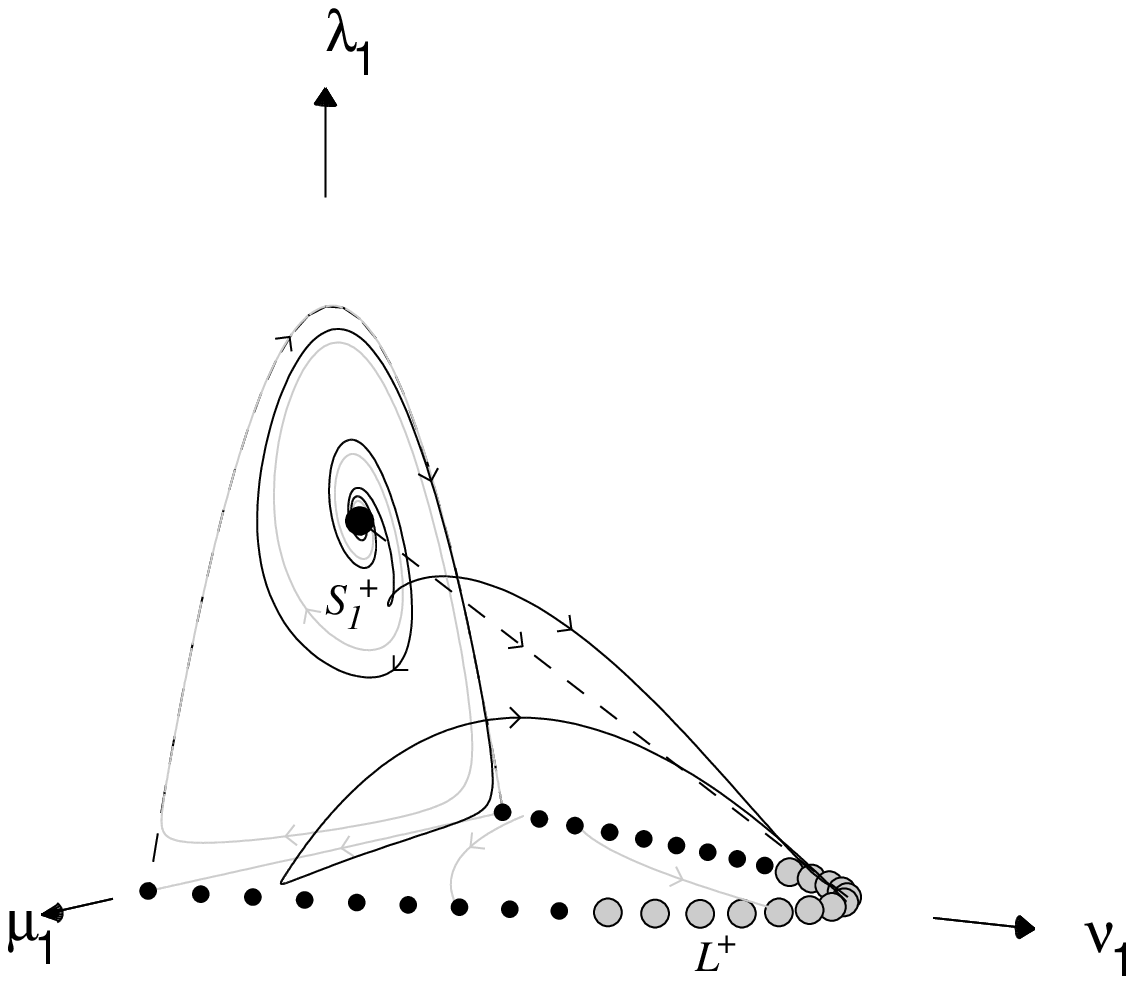}
  \caption{\em Phase portrait of the system
(\ref{dmu_1})-(\ref{dnu_1}) for $\Lambda_{\rm M}>0$
with $\rho\neq0$ and $K=0$.  Note that the label $L^+$ refers to a {\em line}
of equilibrium points.  In this phase space, $\dot\varphi>0$ is
assumed.  See also caption to Fig.  1.}
 \end{figure}

\subsubsection{The Invariant Set $\rho=0$ for $\Lambda_{\rm M}>0$, $K>0$}

For the invariant set $\rho=0$, the system
(\ref{dmu_1})-(\ref{dlambda_1}) reduces to the three dimensions $\{\mu_1,
\chi_1, \nu_1 \}$ ($\lambda_1=1-\mu_1^2-\nu_1$).  
In this invariant set, $\mu$
is a {monotonically decreasing} function, and so the possibility
of periodic orbits is excluded.
The only
equilibrium points are the lines $L^\pm$ with the first
three eigenvalues in Eq. (\ref{Lpm}), and therefore the early and late time
attractors are the lines $L^-$ (for $\mu_1>\frac{-1}{\sqrt{3}}$, 
$h_*>\frac{-1}{3}$) and $L^+$
(for $\mu_1<\frac{1}{\sqrt{3}}$, $h_*<\frac{1}{3}$). The 
curvature and cosmological constant are only dynamically
significant at intermediate times and the phase space is depicted in
Fig.  8.
 \begin{figure}[htp]
  \centering
   \includegraphics*[width=5in]{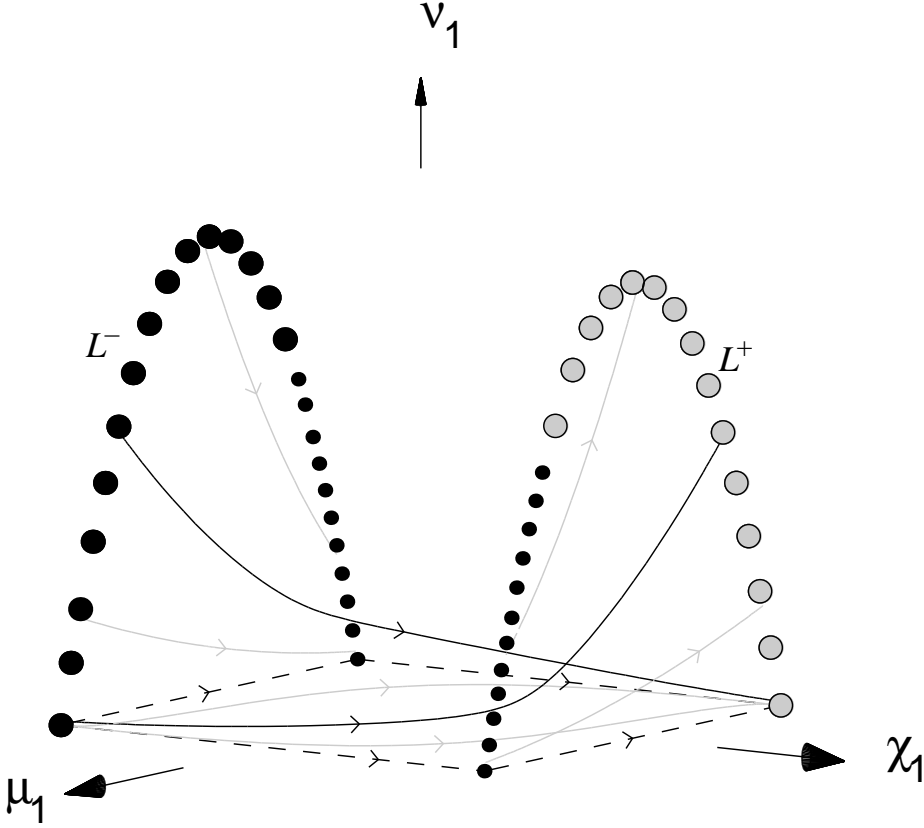}
  \caption{\em Phase portrait of the system
(\ref{dmu_1})-(\ref{dnu_1}) for $\Lambda_{\rm M}> 0$ with
$\rho=0$ field and $K>0$.  Note that $L^+$ and $L^-$ represent {\em lines} of
equilibrium points.  See also caption to Fig.  1.}
 \end{figure}

\subsubsection{Qualitative Analysis of the Four-Dimensional System}

The qualitative dynamics in the full four-dimensional phase space is
as follows. The function
$\chi_1\nu_1^{\frac{1}{2}}\left(1-\mu_1^2-\nu_1-\lambda_1
\right)^{\frac{1}{6}}\lambda_1^{-\third}$ 
monotonically increases and so there can be no periodic orbits.
The only past--attractors belong to the line $L^-$ for
$\mu_1^2< \frac{1}{3}$ $(h_*^2<\frac{1}{9})$
and the only future attractors belong to
the line $L^+$ for $\mu_1^2< \frac{1}{3}$ $(h_*^2< \frac{1}{9})$.  
Both lines $L^\pm$ lie
in both of the invariant sets $\rho=0$ and $\tilde K=0$, which is consistent
with the analysis of Eq. (\ref{dchi}). 
This implies that the spatial curvature, 
cosmological constant and axion field are
only dynamically significant at intermediate times.

For orbits in the four-dimensional phase space, the complex
eigenvalues of the saddle point $S^\pm_1$ suggest that the
heteroclinic sequences may exist in the four--dimensional set.
Indeed, those orbits which asymptote to the $\tilde K=0$ invariant set
{\em do} generically end in a heteroclinic sequence, interpolated
between two equilibrium points representing two
dilaton--vacuum solutions, as discussed in Section 4.1.1.
However, for those orbits which asymptote towards the $\rho=0$
invariant set, there are no heteroclinic sequences (as is evident from
Fig.  8). 

\subsection{The Case $\Lambda_{\rm M}>0$, $K<0$}

For $K<0$, Eq. (\ref{rrfriedmanna}) is written in the new variables as
\begin{equation}
0\leq \mu_2^2+\nu_2+\zeta_2+\lambda_2\leq 1, \qquad \chi_2^2=1,
\end{equation}
where the ``$+$'' sign for $\lambda$ and the ``$-$'' sign for $\zeta$ 
have been chosen in Eq. 
(\ref{TheDefs2}).  For this case, we explicitly choose $\chi_2=+1$, 
as $\chi_2=-1$ corresponds to a time reversal of Eqs. 
(\ref{rr1a})--(\ref{rr5a}).  The system (\ref{rr1a})-(\ref{rr4a}) then
reduces to the four-dimensional system:
\begin{eqnarray}
\label{dmu_2}
\frac{d\mu_2}{d\tau} &=& \left( 1-\mu_2^2-\nu_2-\frac{1}{2}\lambda_2\right)
	\left(\sqrt3+\mu_2\right)-\sqrt3\left(\lambda_2
	+\frac{2}{3}\zeta_2\right) , \\
\label{dnu_2}
\frac{d\nu_2}{d\tau} &=& 2\nu_2\left(1-\mu_2^2-\nu_2
	-\frac{1}{2}\lambda_2\right), \\ 
\label{dzeta_2}
\frac{d\zeta_2}{d\tau} & = & -2\zeta_2\left(\mu_2^2+\nu_2
	+\frac{1}{2}\lambda_2+\frac{1}{\sqrt 3}\mu_2\right), \\
\label{dlambda_2}
\frac{d\lambda_2}{d\tau} &=& \lambda_2\left(1-2\mu_2^2-2\nu_2-\lambda_2
	+\sqrt {3}\mu_2\right).
\end{eqnarray}
The invariant sets $\mu_2^2+\nu_2+\zeta_2+\lambda_2=1$ ($\rho=0$),
$\zeta_2=0$ ($K=0$), $\nu_1=0$ ($N=0$) and $\lambda_1=0$
($\Lambda_{\rm M}=0$) define the boundaries to the phase space.  The
equilibrium points and their respective eigenvalues (denoted by
$\lambda$) are given by
\begin{eqnarray}
S^+: && \mu_2 =-\frac{1}{\sqrt3},\zeta_2=\frac{2}{3} ,\lambda_2=0,\nu_2=0 ; 
	\nonumber \\ 
	&& (\lambda_1,	\lambda_2, \lambda_3,\lambda_4) = 
		\third(2,4,-2,4) \\
N: && \mu_2=-\frac{\sqrt 3}{5}, \nu_2=0,\zeta_2=\frac{18}{25},
	\lambda_2=\frac{4}{25};  \nonumber \\ 
	&& (\lambda_1, \lambda_2, \lambda_3,\lambda_4) = 
	\frac{2}{5}(1+i\sqrt2,1-i\sqrt2, 4, 2) \\
L^+: && \mu_2^2+\nu_2=1, \zeta_2= 0, \lambda_2=0; \nonumber \\ \nonumber
	&&   (\lambda_1, \lambda_2, \lambda_3,\lambda_4) = 
	\left( 0,-\frac{2}{\sqrt3}
	\left[\mu_2+\sqrt3\right],\sqrt3\left[\mu_2-\frac{1}{\sqrt3} 
	\right], -2\sqrt3\left[\mu_2+\frac{1}{\sqrt3}\right]\right) . \\
\end{eqnarray}
{}From the eigenvalues, it is clear that $L^+$ is a late-time
attractor for $\mu_2^2< \frac{1}{3}$ ($h_*^2<\frac{1}{9}$).  The point
$N$ inside the phase space is the early-time attractor for the system,
and represents the curvature--driven solution (\ref{open_new}).  The
saddle point $S^+$ corresponds to the ``$-$'' branch of the Milne
solution (\ref{curv_drive}).

\subsubsection{The Invariant Set $\rho=0$ for $\Lambda_{\rm M}>0$, $K<0$}

For this invariant set, the four-dimensional system
(\ref{dmu_2})-(\ref{dlambda_2}) reduces to a three-dimensional system
involving the coordinates $\{\mu_2,\nu_2,\lambda_2\}$
($\zeta_2=1-\mu_2^2-\nu_2-\lambda_2$).  The equilibrium points are the
same as for the full four-dimensional set, but the eigenvalues are now
($\lambda_1,\lambda_2,\lambda_3$). The variable $\nu$ is a
monotonically increasing function, the existence of which eliminates
the possibility of recurrent orbits.  Thus, the generic behaviour of
this model is for solutions to asymptote into the past towards the
curvature--dominated solution (\ref{open_new}), represented by the
point $N$, and to the future towards the line $L^+$ for $\mu_2<
\frac{1}{\sqrt{3}}$ ($h_*< \frac{1}{3}$). Fig.  9 depicts this phase
space.
 \begin{figure}[htp]
  \centering
   \includegraphics*[width=5in]{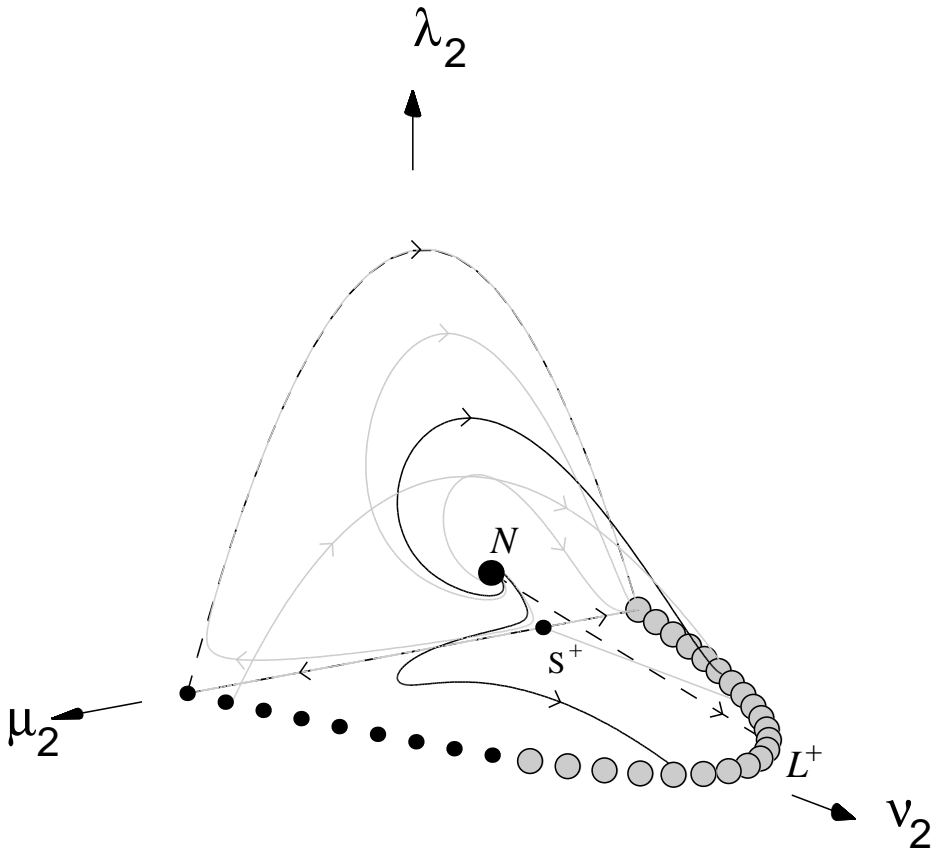}
  \caption{\em Phase portrait of the system
(\ref{dmu_2})-(\ref{dnu_2}) for $\Lambda_{\rm M}>0$
with $\rho=0$ and $K<0$.  Note that $L^+$ represents a {\em line} of
equilibrium points.  In this phase space, $\dot\varphi>0$ is assumed.
See also caption to Fig.  1.}
 \end{figure}

\subsubsection{Qualitative Analysis of the Four-Dimensional System}

The invariant set $K=0$ was described in Section 4.1.1 (see Fig. 7).
In the full four-dimensional set, the point $N$ is the early-time
attractor, and $L^+$ is the late-time attractor for
$\mu_2^2<\frac{1}{3}$ ($h_*^2<\frac{1}{9}$).  The point $N$ lies in the invariant set
$\rho=0$ and $L^+$ lies in both of the invariant sets $\rho=0$ and
$\tilde K=0$. This is consistent with the analysis of
Eq. (\ref{dchi}).  We see that $\nu_2$ is a monotonically
increasing function, and so there are no recurrent or periodic
orbits.  Furthermore, since $\nu_2$ increases  
monotonically, the shear in the model is initially dynamically
trivial, but becomes significant asymptotically into the
future.  The axion field and cosmological constant are only dynamically
important at intermediate times and do not play a r\^{o}le in
the early- and late-time behaviour of the cosmologies.  The curvature term is
dynamically significant at early and intermediate times, but becomes
dynamically trivial at late times.

Orbits which asymptote into the future towards the $K=0$ invariant
set generically end in a heteroclinic sequence,  as described in
Section 4.1.1 and depicted in Fig. 7.  However, such a sequence does
not occur for orbits which asymptote into the future towards the
$\rho=0$ invariant set.  Indeed, by examining the eigenvalues of the
equilibrium points of the four-dimensional system, there do not seem
to be heteroclinic sequences outside of the $\tilde K=0$ invariant
set.

\subsection{The Case $\Lambda_{\rm M}<0$, $K>0$}

For completeness we 
now consider the cases where $\Lambda_{\rm M} < 0$. When $K>0$, 
Eq. (\ref{rrfriedmanna}) is written in terms of the new variables as
\begin{equation}
0\leq \mu_3^2+\nu_3\leq 1, \qquad \chi_3^2+\zeta_3+\lambda_3=1,
\end{equation}
where the ``$-$'' sign for $\lambda$ and the ``$+$'' sign for $\zeta$
have been chosen in Eq. (\ref{TheDefs2}).  The variable $\lambda_3$ is
chosen as the extraneous variable and the system
(\ref{rr1a})-(\ref{rr4a}) then reduces to the four-dimensional system:
\begin{eqnarray}
\label{dmu_3}
\frac{d\mu_3}{d\tau} &=& \left(1-\mu_3^2-\nu_3\right)\left(\sqrt3
	+\mu_3\chi_3\right) +\frac{\sqrt 3}{2}\left(1-\mu_3^2\right)
	\left(1-\chi_3^2-\frac{5}{3}\zeta_3\right), \\
\nonumber
\frac{d\chi_3}{d\tau} &=& -\frac{\sqrt3}{2}\left[ \mu_3\chi_3\left( 1-\chi_3^2
	-\frac{5}{3}\zeta_3\right)\right] -\frac{1}{2}\left(1-\chi_3^2\right)
	\left(1-2\mu_3^2-2\nu_3\right)+\frac{1}{2}\zeta_3 \\ \label{dchi_3} \\
\label{dnu_3}
\frac{d\nu_3}{d\tau} &=&\nu_3\left[ 2\chi_3\left(1-\mu_3^2-\nu_3\right)-\sqrt 3
	\mu_3\left(1-\chi_3^2-\frac{5}{3}\zeta_3\right)\right], \\
\label{dzeta_3}
\frac{d\zeta_3}{d\tau} &=& -\zeta_3 \left[ 2\chi_3\left(\mu_3^2+\nu_3\right) 
	+\frac{1}{\sqrt 3}\mu_3\left(5-3\chi_3^2-5\zeta_3\right)\right].
\end{eqnarray}
The invariant sets $\mu_3^2+\nu_3=1$, $\chi_3^2+\zeta_3=1$, $\nu_3=0$
and $\zeta_3=0$ define the boundary to the phase space.  The
equilibrium sets and their corresponding eigenvalues (denoted by
$\lambda$) are
\begin{eqnarray}
\nonumber
L^\pm : && \chi_3=\pm 1, \mu_3^2+\nu_3=1, \zeta_3=0; \\ \nonumber
	&& (\lambda_1,\lambda_2,\lambda_3,\lambda_4) = \left( 0, \mp\frac{2}
	{\sqrt3}\left[\sqrt3\pm\mu_3\right], \sqrt{3}\left[\mu_3\mp\frac{1}
	{\sqrt3}\right], -2\sqrt 3\left[\mu_3\pm\frac{1}{\sqrt3} 
	\right]\right),\\ \\
\nonumber
R: && \chi_3=-\frac{1}{\sqrt3}, \mu_3=-1, \nu_3=0, \zeta_3=0; \\
   && (\lambda_1,\lambda_2,\lambda_3,\lambda_4) = \frac{1}{\sqrt3}
	\left(1, 2, 6, 10\right), \\
\nonumber
A: && \chi_3=\frac{1}{\sqrt3}, \mu_3=1, \nu_3=0, \zeta_3=0; \\
   && (\lambda_1,\lambda_2,\lambda_3,\lambda_4) = -\frac{1}{\sqrt3}
	\left(1, 2, 6, 10\right).
\end{eqnarray}
Here, there are two early-time attractors. The first is the 
point $R$,  representing
the ``$+$'' branch of the constant axion, spatially isotropic  solution 
(\ref{at_half}), where $\dot\varphi<0$. The second is the line
$L^-$ for $\mu_3^2< \frac{1}{3}$ ($h_*^2<\frac{1}{9}$).
Likewise, there are two late-time attractors: these are the point $A$, 
representing the  ``$-$'' solution of Eq. (\ref{at_half}), where 
$\dot\varphi>0$, and the line $L^+$ for $\mu_3^2<\frac{1}{3}$ ($h_*^2<\frac{1}{9}$).

\subsubsection{The Invariant Set $K=0$ for $\Lambda_{\rm M}<0$ \label{K0LRn}}

In paper II, this invariant set was examined using
variables which are the same as $\{\mu_3,\chi_3,\nu_3\}$.  There it 
was shown that $\mu_3$ is {monotonically increasing}, so that most
trajectories in this phase space represent bouncing cosmologies
which are initially contracting.  Most
trajectories asymptote into the past towards either $L^-$ 
or to $R$ (see above). To the future, most orbits
asymptote towards either $L^+$ or $A$.  Fig.  10 depicts this
three-dimensional phase space.
 \begin{figure}[htp]
  \centering
   \includegraphics*[width=4in]{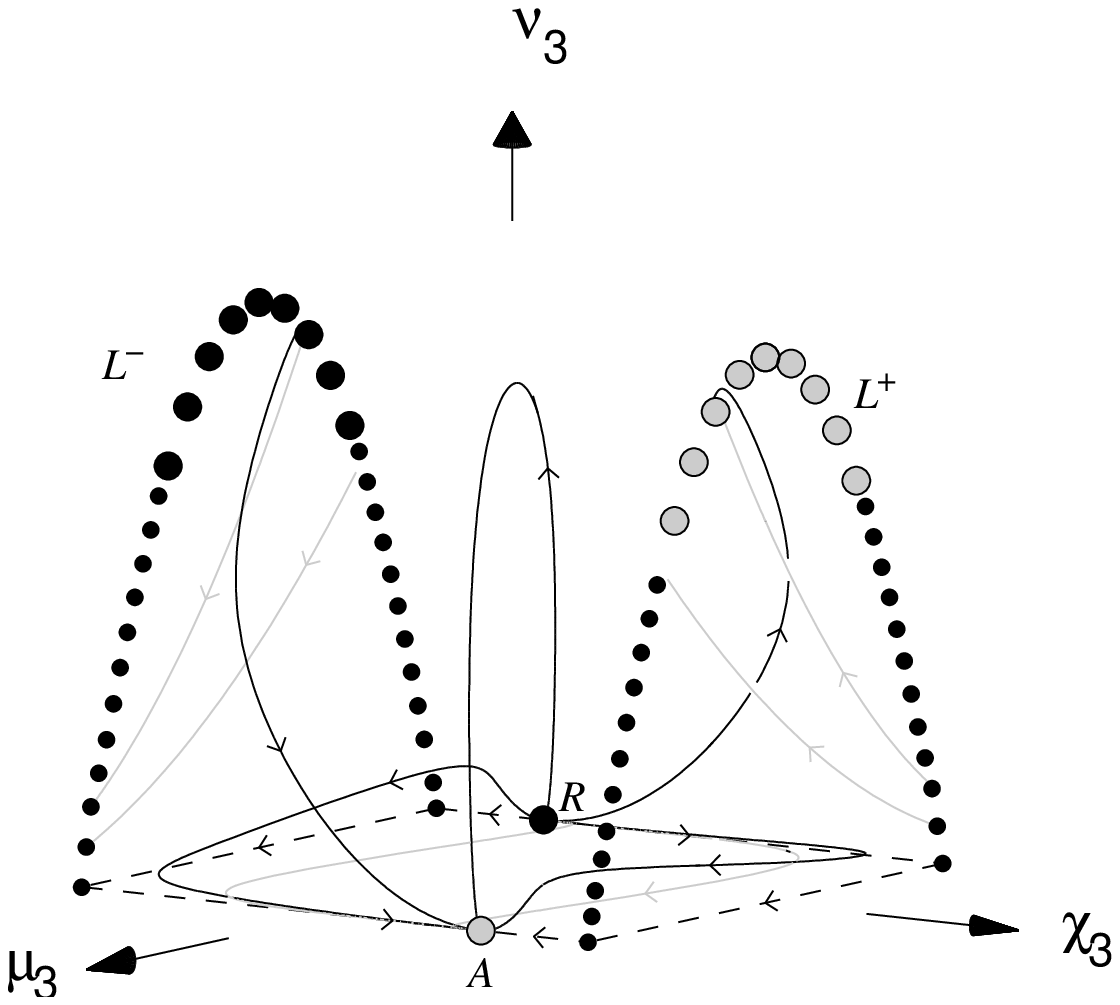}
  \caption{\em Phase portrait of the system
(\ref{dmu_3})-(\ref{dzeta_3}) for $\Lambda_{\rm M}<0$
with $\rho\neq0$ and $K=0$.  Note that the labels $L^+$ and $L^-$ refer to 
{\em lines} of equilibrium points.  See also caption to Fig.  1.}
 \end{figure}

\subsubsection{The Invariant Set $\rho=0$ for $\Lambda_{\rm M}<0$, $K>0$}

For this invariant set, the four-dimensional system
(\ref{dmu_3})-(\ref{dzeta_3}) reduces to a three-dimensional system
involving the coordinates $\{\mu_3,\chi_3,\zeta_3\}$
($\nu_3=1-\mu_2^3$).  The equilibrium points are the same as the full
four-dimensional set, but with eigenvalues
($\lambda_1,\lambda_2,\lambda_3$), and so the line $L^-$ is a source
for $\mu_3>\frac{-1}{\sqrt{3}}$ ($h_*>\frac{-1}{3}$)
and the line $L^+$ is a sink for
$\mu_3<\frac{1}{\sqrt{3}}$ ($h_*< \frac{1}{3}$). 
The function $\chi_3/\sqrt{\nu_3}$ is 
monotonically increasing, and so there are no recurring or periodic
orbits.  Hence, solutions generically asymptote into the past towards
 $L^-$ or $R$ and into the future towards $L^+$ or $A$. 
 \begin{figure}[htp]
  \centering
   \includegraphics*[width=5in]{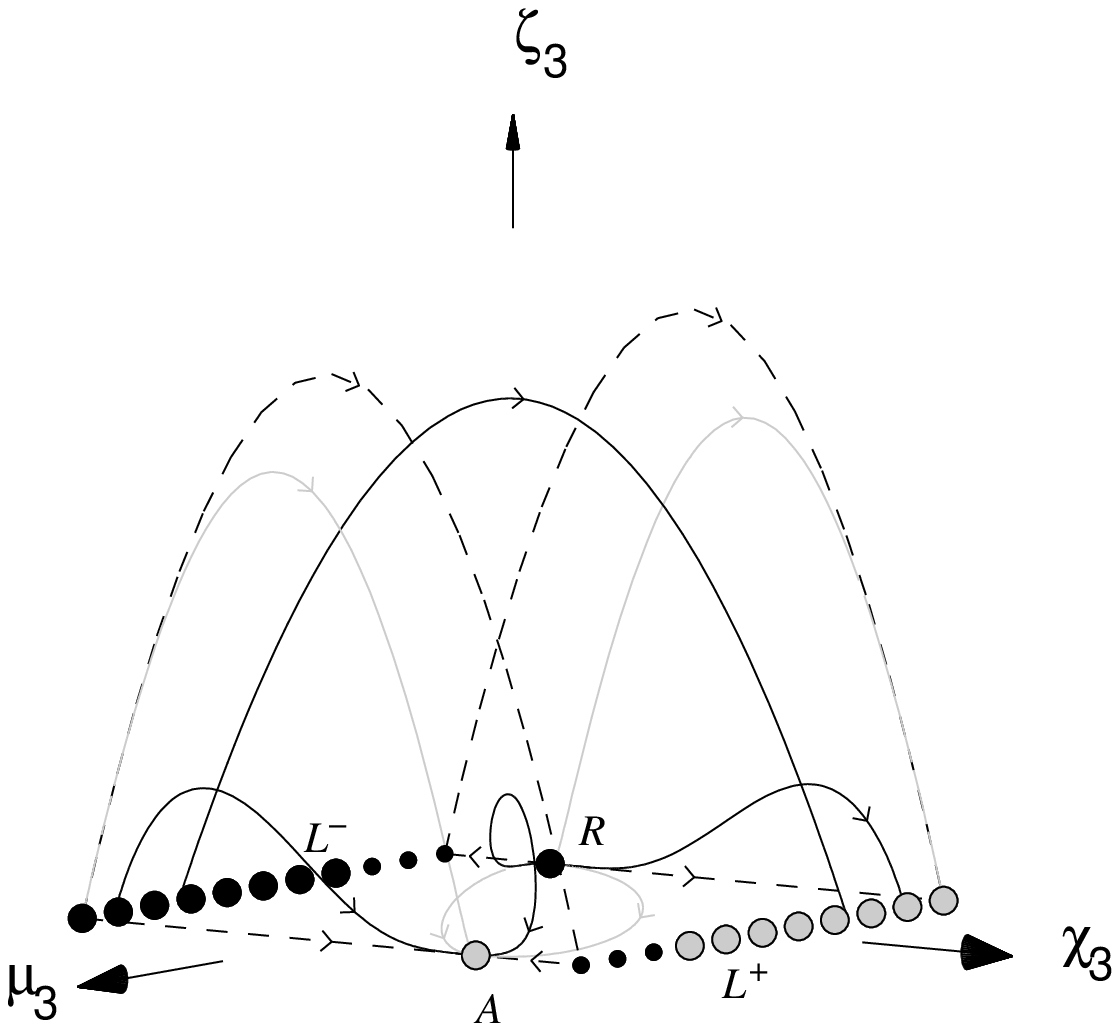}
  \caption{\em Phase portrait of the system
(\ref{dmu_3})-(\ref{dzeta_3}) for $\Lambda_{\rm M}<0$
with $\rho=0$ and $K>0$.  Note that $L^+$ and $L^-$ represent
{\em lines} of equilibrium points.  See also caption to Fig.  1.}
 \end{figure}

\subsubsection{Qualitative Analysis of the Four-Dimensional System}

The function $\left(\chi_3+\frac{1}{\sqrt3}\mu_3\right)/\sqrt{\nu_3}$
is {monotonically increasing}, and so there are no recurring or
periodic orbits in 
the full four-dimensional set. The early-time attractors are the
line $L^-$ for $\mu_3^2< \frac{1}{3}$ 
($h_*^2<\frac{1}{9}$) and the point $R$.  The late-time
attractors are the line $L^+$ for $\mu_3^2< \frac{1}{3}$ ($h_*^2<\frac{1}{9}$) 
and the point $A$.  All
of these attractors lie in both of the invariant sets $\rho=0$ and
$\tilde K=0$  and the axion field and curvature term are
dynamically significant only at intermediate times.  For early and
late times, the cosmological constant is dynamically important only
when the shear is dynamically trivial (e.g., the points $R$
and $A$), and vice-versa (e.g., the lines $L^\pm$).

\subsection{The Case $\Lambda_{\rm M}<0$, $K<0$}

For this case, Eq. (\ref{rrfriedmanna}) is written in 
terms of the new variables as
\begin{equation}
0\leq \mu_4^2+\nu_4+\zeta_4\leq 1, \qquad \chi_4^2+\lambda_4=1,
\end{equation}
where the ``$-$'' sign for both $\lambda$ and $\zeta$
has been chosen in Eq. (\ref{TheDefs2}).  The variable $\lambda_4$ is
chosen as the extraneous variable and the system
(\ref{rr1a})-(\ref{rr4a}) then reduces to the four-dimensional system:
\begin{eqnarray}
\label{dmu_4}
\frac{d\mu_4}{d\tau} &=& \left(1-\mu_4^2-\nu_4\right)\left(\sqrt 3+\mu_4\chi_4
\right)+\frac{\sqrt3}{2}\left(1-\mu_4^2\right)\left(1-\chi_4^2\right)
-\frac{2}{\sqrt3}\zeta_4, \\
\frac{d\chi_4}{d\tau} &=& -\frac{1}{2}\left(1-\chi_4^2\right)\left[ 
	1-2\mu_4^2-2\nu_4+\sqrt3 \mu_4\chi_4\right], \\
\label{dnu_4}
\frac{d\nu_4}{d\tau} &=&\nu_4\left[ 2\chi_4\left(1-\mu_4^2-\nu_4\right)-\sqrt 3
	\mu_4\left(1-\chi_4^2\right)\right], \\
\label{dzeta_4}
\frac{d\zeta_4}{d\tau} &=& -\zeta_4 \left[ 2\chi_4\left(\mu_4^2+\nu_4\right) 
	+\frac{1}{\sqrt 3}\mu_4\left(5-3\chi_4^2\right)\right].
\end{eqnarray}
The invariant sets $\mu_4^2+\nu_4+\zeta_4=1$, $\chi_4^2=1$, $\nu_4=0$
and $\zeta_4=0$ define the boundary to the phase space.  The
equilibrium sets and their corresponding eigenvalues (denoted by
$\lambda$) are
\begin{eqnarray}
\nonumber
L^\pm : && \chi_4=\pm 1, \mu_4^2+\nu_4=1, \zeta_4=0; \\ \nonumber
	&& (\lambda_1,\lambda_2,\lambda_3,\lambda_4) = \left( 0, \mp\frac{2}
	{\sqrt3}\left[\sqrt3\pm\mu_4\right], \sqrt{3}\left[\mu_4\mp \frac{1}
	{\sqrt3}\right], -2\sqrt 3\left[\mu_4\pm\frac{1}{\sqrt3} 
	\right]\right),\\ \\
\nonumber
R: && \chi_4=-\frac{1}{\sqrt3}, \mu_4=-1, \nu_4=0, \zeta_4=0; \\
   && (\lambda_1,\lambda_2,\lambda_3,\lambda_4) = \frac{1}{\sqrt3}
	\left(1, 2, 6, 10\right), \\
\nonumber
A: && \chi_4=\frac{1}{\sqrt3}, \mu_4=1, \nu_4=0, \zeta_4=0; \\
   && (\lambda_1,\lambda_2,\lambda_3,\lambda_4) = -\frac{1}{\sqrt3}
	\left(1, 2, 6, 10\right),\\
\nonumber
S^\pm : && \chi_4=\pm 1,\mu_4=\frac{\mp 1}{\sqrt3},\nu_4=0,\zeta_4=\frac{2}{3}; \\
      && (\lambda_1,\lambda_2,\lambda_3,\lambda_4) = \frac{2}{3} 
         (\mp 1, \pm 1, \pm 2, 0).
\end{eqnarray}
As in the previous case, there are two early-time attractors: the
point $R$ and
the line $L^-$ for $\mu_4^2<\frac{1}{3}$ ($h_*^2<\frac{1}{9}$).
Likewise, there are two late-time attractors: the point $A$
and the line $L^+$ for $\mu_4^2<\frac{1}{3}$ ($h_*^2<\frac{1}{9}$).

\subsubsection{The Invariant Set $\rho=0$ for $\Lambda_{\rm M}<0$, $K<0$}

For this invariant set, the four-dimensional system
(\ref{dmu_4})-(\ref{dzeta_4}) reduces to a three-dimensional system
involving the coordinates $\{\mu_4,\chi_4,\nu_4\}$
($\zeta_4=1-\mu_4^3-\nu_4$).  The equilibrium points are the same as
the full four-dimensional set,  and the eigenvalues are 
($\lambda_1,\lambda_2,\lambda_3$). The difference is 
that the line $L^+$ is a sink
for $\mu_4<\frac{1}{\sqrt{3}}$ and the line $L^-$ is a source for
$\mu_4>\frac{-1}{\sqrt{3}}$.  The function
$\mu_4/\sqrt{\nu_4}$ is {monotonically
increasing}, and so periodic orbits cannot occur.
Hence, solutions generically asymptote into the past towards
the ``$+$'' branch of 
Eq. (\ref{dmv}) for $h_*>\frac{-1}{3}$, 
or the ``$+$'' solution of Eq. (\ref{at_half}). Into
the future, solutions asymptote towards either the ``$-$'' 
branch of Eq. (\ref{dmv}) for $h_*<\frac{1}{3}$, or to the ``$-$''
solution of Eq. (\ref{at_half}). Fig.  12
depicts this phase space.
 \begin{figure}[htp]
  \centering
   \includegraphics*[width=5in]{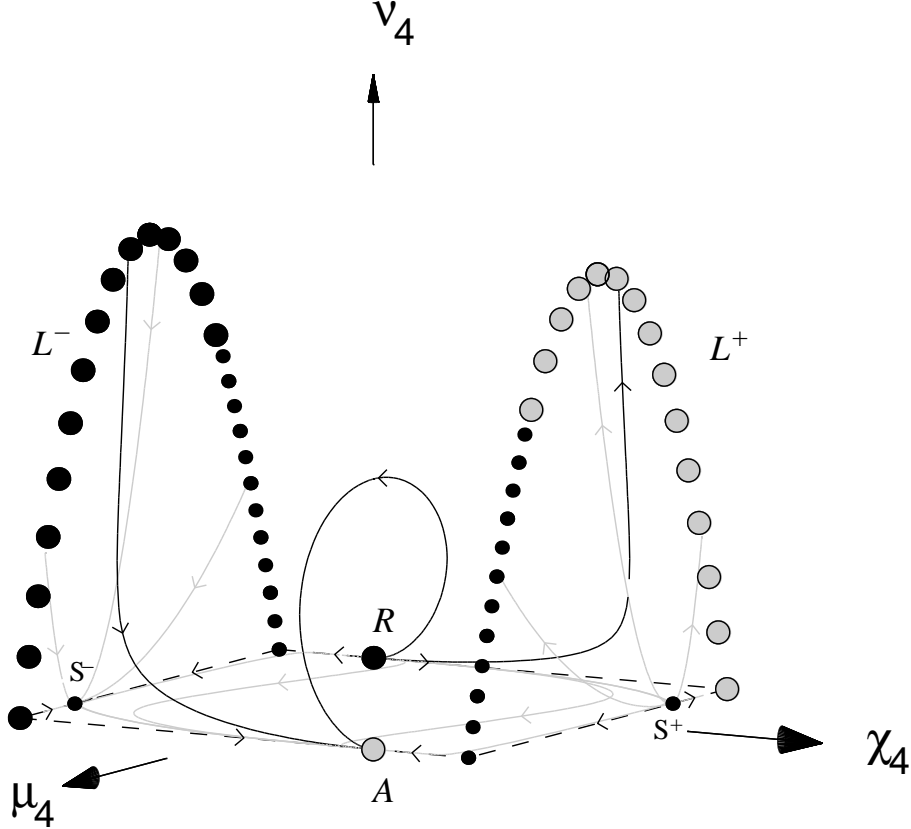}
  \caption{\em Phase portrait of the system
(\ref{dmu_4})-(\ref{dzeta_4}) for $\Lambda_{\rm M}<0$
with $\rho=0$ and $K<0$.  Note that $L^+$ and $L^-$ represent
{\em line} of equilibrium points. See also caption to Fig. 1.}
 \end{figure}

\subsubsection{Qualitative Analysis of the Four-Dimensional System}

The invariant set $K=0$ is discussed in Section 4.3.1 (see Fig. 10).
The function $\mu_4/\sqrt{\nu_4}$ is {monotonically increasing},
and so there are no recurring or periodic orbits.
The asymptotic behaviour is identical to that described in 
Section 4.3.3. 

\setcounter{equation}{0}
\section{Discussion}

In this paper, we have presented a complete qualitative analysis for
the isotropic curvature string cosmologies derived from the 
effective action (\ref{singleaction}) in the two cases where either
$\Lambda =0$ or $\Lambda_{\rm M} =0$ (the cases where both terms are
non-zero is discussed in \cite{BillyardThesis}). This was made
possible by compactifying the phase space in terms of suitably defined
variables.  When $\Lambda_{\rm M} =0$, Eq. (\ref{singleaction})
represents the action for the NS--NS fields that arise in both the
type II and heterotic string theories when an arbitrary central charge
deficit is present.  We identified the cosmological constant
$\Lambda_{\rm M}$ with terms that arise in the RR sector of the
massive type IIA supergravity theory when appropriate conditions 
apply. 

The subset $\dot{\beta}=0$ corresponds to the 
class of spatially isotropic FRW universes with arbitrary 
spatial curvature. In the positively curved case, we 
have extended the work of Easter {\em et al.}
\cite{emw}, who performed a perturbation analysis on the static, closed
FRW model to show that it was a late-time attractor.
More generally, the models we have considered 
represent Bianchi type I, V and IX universes. 

For each case,  we
have established the existence 
of monotonic functions which precludes the
existence of recurrent or periodic orbits. Consequently, 
the early-- and late--time behaviour of these models can be determined 
by analysing the nature of the equilibrium points/lines of the system. 
In all cases, the spatially flat 
dilaton--vacuum solutions $L^{\pm}$, given by Eq. (\ref{dmv}), 
act as either early- or
late-time attractors and, in many cases, act as both.  Because these
solutions lie in both the $\rho=0$ and $\tilde K=0$ invariant sets and
contain neither a central charge deficit nor a non--zero $\Lambda_{\rm M}$ 
contribution, 
we may conclude that the shear and dilaton fields are dynamically
dominant asymptotically.  Furthermore, with the exception of the
$\Lambda>0$, $\tilde K>0$ case, all early-time and late-time
attracting sets lie in either the $\rho=0$ invariant set or the
$\tilde K=0$ invariant set, and a majority of these sets lie in both.

Thus, we see a generic feature in which the curvature terms and the
axion field are dynamically significant at intermediate times and are
asymptotically negligible at early and late times.  The {\em
exception} to this generic behaviour is the $\Lambda>0$, $\tilde K>0$
case, where the generalized linear dilaton--vacuum solution
(\ref{static_general}), in which neither $\rho=0$ nor $\tilde K=0$,
acts both as a repeller (for $\dot\varphi<0$) and as an attractor (for
$\dot\varphi>0$).  In these solutions, the variables $\rho$ and
$\tilde K$ are proportional to the central charge deficit, $\Lambda$.
Note that asymptotically $\dot\alpha=0$ (and $\dot\beta=0$) and hence
these models are static.

When $\Lambda<0$, the central charge deficit is dynamically
significant only at intermediate times, and is asymptotically
negligible at early and late times.  In fact, the only repelling and
attracting sets in this instance are the dilaton--vacuum
solutions.  When $\Lambda>0$, the central charge deficit can be
dynamically significant at both early and late times, and the corresponding
solution is the generalized linear dilaton--vacuum solution
(\ref{static_general}), represented by the line $L_1$.  When $\tilde
K>0$, these solutions can be repelling ($\dot\varphi<0$) and
attracting ($\dot\varphi>0$).  When $\tilde K<0$, the endpoint of
this line, $C$ (representing Eq. (\ref{static})) is a repeller.

When $\Lambda_{\rm M}>0$, the cosmological constant may play a
significant r\^{o}le in the early and late time dynamics.  For
instance, although in the four-dimensional sets there are no repelling
or attracting equilibrium points in which $\Lambda_{\rm M}$ is
dynamically significant, we have found that the orbits which are
attracted to the $\tilde K=0$ invariant set end in a heteroclinic
sequence which interpolates between two dilaton--vacuum
solutions (see Section 4.1.1, Fig. 7 and paper II).
During this interpolation, the orbits repeatedly spend time in a
region of phase space in which $\Lambda_{\rm M}$ is dynamically
significant (the region $\lambda_1>0$ in Fig. 7), although most time
is spent near the dilaton--vacuum saddle points where
$\Lambda_{\rm M}$ is dynamically negligible.  When $\Lambda_{\rm
M}<0$, the cosmological constant can be dynamically significant at
both early and late times, since solutions typically
asymptote to the solution (\ref{at_half}), where the shear and axion field 
are static. For
the repelling and attracting sets of the $\Lambda_{\rm M}<0$ cases,
the shear term is only dynamically significant when the
cosmological constant is {\em not}, and vice versa.

The parameter $\beta$ measures the degree of anisotropy in the models.
If we define isotropization by the condition $\dot\beta\rightarrow 0$
\cite{WE,Collins1973a}, then we note that in general $\dot\beta\neq 0$
at the equilibrium points on the line $L^\pm$.  Therefore, solutions
asymptoting towards the sinks on these lines do not isotropize to the
future.  At all other equilibrium points, however, $\dot\beta=0$ and
the {\em corresponding appropriate string cosmologies therefore
`isotropize'}.  This is an important result and a similar situation
occurs in the time--reverse models.  On the other hand, the question
of isotropization of string cosmologies in a more general context
remains an open question.  Note that the shear in the models that we
have discussed is essentially of Bianchi type I.  In general
relativity with a perfect fluid, it is known that Bianchi type I
models isotropize whereas in general spatially--homogeneous models do
{\em not} isotropize \cite{WE,Collins1973a}).

For the equilibrium points corresponding to the sinks on the line
$L^+$, $\dot\beta^2\propto\dot\varphi^2$ ($\rho=\tilde K=0$), and so
the energy densities of the modulus and dilaton fields are
proportional to one another.  
Hence, the corresponding dilaton--vacuum solutions are
``matter scaling'' string cosmology solutions, which act as local
attractors, similar to the matter scaling solutions in general relativistic
scalar field cosmologies \cite{Billyard1999e_Copeland1997a}.
Finally, for every equilibrium point within these phase spaces, the
scale factor of the corresponding exact solution is a power--law
function of cosmic time, and therefore all of the corresponding exact
solutions are self--similar \cite{Maartens}.

The cases in which $\dot\varphi<0$ are related to a time reversal of
the models discussed in the text.  We have not explicitly considered
these models here; however, the conclusions concerning isotropization
are similar although the details of these models may be different
(e.g., the r\^{o}le of sources and sinks can be interchanged).

In conclusion, therefore, we have established the qualitative properties of 
all the isotropic curvature string 
models discussed in the text by finding appropriate monotone 
functions. Typically, {\em the curvature term is dynamically
significant only at intermediate times} and is asymptotically
negligible.  There are only two exceptions to this. The 
first corresponds to the case
$\{\Lambda>0 , \tilde K>0 \}$, where the generalized linear dilaton--vacuum
attractors and repellers have a non-negligible curvature. The 
second case is the 
$\{ \Lambda_{\rm M}>0, \tilde K<0 \}$ model in which the repeller $N$
represents the curvature--driven 
solution (\ref{open_new}).  Finally, we note that when 
$\Lambda_{\rm M}>0$ there exist heteroclinic sequences in the
invariant set $\tilde K=0$. This implies that the qualitative
behaviour associated with heteroclinic sequences is only
valid for solutions which approach $\tilde K=0$.

\centerline{\bf Acknowledgments}

\vspace{.3in}
APB is supported by Dalhousie University, AAC is supported
by the Natural Sciences and Engineering Research Council
of Canada (NSERC), and
JEL is supported by the Royal Society. 

\vspace{.7in}

\appendix
\renewcommand\theequation{A.\arabic{equation}}
\setcounter{equation}{0}

\section{Equilibrium Points}

In this Appendix, we present the analytical solutions 
to Eqs. (\ref{rr1})--(\ref{rrfriedmann}) that represent 
all of the equilibrium points that arise in this paper.  

The `dilaton-vacuum' solutions correspond to solutions 
where the axion field is constant and cosmological constants vanish 
$(\Lambda=\Lambda_{\rm M}=\dot{\sigma} =0)$. In the spatially 
flat case $(\tilde{K} =0)$, they are power laws: 
\begin{eqnarray}
\nonumber
a & = & a_*\left| t \right|^{\pm h_*}, \nonumber \\
\nonumber
e^\Phi & = &e^{\Phi_*}\left|t\right|^{\pm3h_*-1}, \nonumber \\
\nonumber
e^\beta & =& e^{\beta_*}\left|t\right|^{\pm\epsilon\sqrt{(1-3h_*^2)/6}} ,\\
\nonumber
\sigma&=&\sigma_*,\\
\label{dmv}
k&=&0,
\end{eqnarray}
where $\{ a_* , \Phi_* ,\sigma_*, \beta_* ,h_*\}$ are constants, $a\equiv 
e^{\alpha}$ is the averaged scale
factor of the universe, the $\pm$ sign corresponds to the
sign of $t$ and $\epsilon=\pm 1$. These solutions have a curvature 
singularity at $t=0$. 
Note that the shifted dilaton field (\ref{varphi}) satisfies 
$\dot\varphi >0$ for $t<0$ and $\dot\varphi<0$
for $t>0$.  The ``$-$'' branch of Eq. (\ref{dmv}) 
corresponds to the line $L^+$
throughout this paper and the ``$+$'' branch corresponds to the line
$L^-$.

Another solution which appears when both 
$\Lambda=\Lambda_{\rm M}=0$ and $\dot{\sigma} =0$ 
is the Milne form of flat space: 
\begin{eqnarray}
\nonumber
a&=&a_*\left(\pm t\right), \\
\nonumber
\Phi&=&\Phi_*,\\
\nonumber
\beta&=&\beta_*,\\
\nonumber
\sigma&=&\sigma_*,\\
\label{curv_drive}
k&=&-a_*^2,
\end{eqnarray}
where $\{a_*,\Phi_*,\beta_*,\sigma_*\}$ are constants.  The ``$\pm$''
sign corresponds to the sign of $t$.  These solutions are labelled
$S^\pm$ throughout the paper (note $S^+$ corresponds to $t<0$ and
$S^-$ corresponds to $t>0$) and arise as saddle points.

There is a line of equilibrium points that arises when the central charge 
deficit, $\Lambda$, is included in the action (\ref{singleaction}). 
This class of solutions has the form 
\begin{eqnarray}
\nonumber
a&=&a_*, \\
\nonumber
\Phi&=&\Phi_* +n\sqrt{\frac{6\Lambda}{2+n^2}} t, \\
\nonumber
\beta&=&\beta_*, \\
\nonumber
\sigma&=&\sigma_* \pm \sqrt{\frac{2\left(1-n^2\right)}{3n^2}} \exp
\left(-\Phi_* -n\sqrt{\frac{6\Lambda}{2+n^2}} t \right), \\
\label{static_general}
k&=&\frac{1-n^2}{2+n^2}\Lambda a_*^2, 
\end{eqnarray}
where $n\in [-1,1]$.  The solution is static, but has non--trivial 
spatial curvature, dilaton and axion fields. 
These solutions are represented by the line $L_1$ in
Section 3.1. In that Section they represent 
past attractors for $n\in(0,1]$ and future attractors for
$n\in[-1,0)$. These solutions were found in \cite{emw}
for $\dot\varphi>0$, where, by employing a 
perturbation analysis, they were found to be late-time attractors.

The endpoints of the line $L_1$ correspond to 
$n=\pm 1$. These represent spatially flat and isotropic solutions known as the 
`linear dilaton--vacuum' solutions \cite{myers}. 
They are described by
\begin{eqnarray}
\nonumber
a&=&a_*, \\
\nonumber
\Phi&=&\Phi_*\pm\sqrt{2\Lambda} t,\\
\nonumber
\beta&=&\beta_*, \\
\nonumber
\sigma&=&\sigma_*,\\
\label{static}
k&=&0,
\end{eqnarray}
where $\{a_*, \Phi_*, \beta_*,\sigma_*\}$ are constants.  The ``$+$''
solution is represented by the equilibrium point $C$ in Section 3.2.

We now consider the cosmologies where $\Lambda=0$ in the action 
(\ref{singleaction}). 
A spatially flat solution found previously in paper II is given by 
\begin{eqnarray}
\nonumber
a & = &  a_* \left[\frac{\sqrt{3\Lambda_{\rm M}}}{4} |t|\right]^{\third} \\
\nonumber
\Phi & = &  \Phi_*-\ln \left[\frac{3\Lambda_{\rm M}}{16} t^2 \right] \\
\nonumber
\beta& = & \beta_*,\\
\nonumber
\sigma & = & \sigma_*\pm \frac{\sqrt{15}}{16}\Lambda_{\rm M} t^2 \\
\label{newsol}
k&=&0,
\end{eqnarray}
where $\{ a_*, \Phi_* ,\beta_* , \sigma_*\}$ are arbitrary constants
and time is defined over the interval $t<0$ ($\dot\varphi>0$) for the
equilibrium point $S^+_1$ and $t>0$ ($\dot\varphi<0$) for the
equilibrium point $S^-_1$ in Section 4.1. 

There also exists a spatially curved, isotropic solution with 
a constant axion field. It is given by
\begin{eqnarray}
\nonumber
a &=& \frac{1}{2}a_*\sqrt \Lambda_{\rm M}\left| t\right|, \\
\nonumber
\Phi &=& - \ln \left[ \frac{1}{4}\Lambda_{\rm M} t^2 \right], \\
\nonumber
\beta &=& \beta_* \\
\nonumber 
\sigma&=&\sigma_*, \\
\label{open_new}
k &=& -\frac{3}{4}\Lambda_{\rm M} a_*^2 
\end{eqnarray}
where $\{ a_*, \beta_* ,\sigma_* \}$ are
integration constants. The solution for 
$t<0$ is represented by the equilibrium point $N$ in
Section 4.2.

For negative $\Lambda_{\rm M}$, there are also the solutions
\begin{eqnarray}
\nonumber
a & = & \frac{a_*}{\sqrt{\pm2t}}, \\
\nonumber
\Phi & = & \Phi_* -\ln\left[-2\Lambda_{\rm M} t^2 \right],  \\
\nonumber
\beta&=&\beta_*, \\
\nonumber
\sigma&=&\sigma_*,\\
\label{at_half}
k&=&0,
\end{eqnarray}
where $\{a_*,\Phi_*,\beta_*,\sigma_*\}$ are constants.  The $\pm$ sign
corresponds to the the sign of $t$ and ``$+$'' solution corresponds to the repelling equilibrium
point $R$ in Section 4.4, whereas the ``$-$'' solution 
corresponds to the attracting equilibrium point $A$ in that Section.

\end{document}